\documentclass[a4paper,11pt]{article}
\usepackage{jcappub} 
\usepackage{lineno}

\usepackage{dcolumn}
\usepackage{verbatim}

\usepackage{breakurl}

\usepackage{lmodern}

\pdfoutput=1
\usepackage{graphicx}
\usepackage{bm}
\usepackage{amsmath}
\usepackage{amsfonts}
\usepackage{amssymb}
\usepackage{multirow}
\usepackage[capitalise]{cleveref}
\usepackage{acro}
\usepackage[utf8]{inputenc}

\usepackage{tikz}
\usepackage{pgfplots}

\crefname{figure}{Fig.}{Figs.}

\newcommand{\be}{{\begin{eqnarray}}}
\newcommand{\ee}{{\end{eqnarray}}}

\newcommand{\overbar}[1]{\mkern 1.5mu\overline{\mkern-1.5mu#1\mkern-1.5mu}\mkern 1.5mu}

\newcommand{\Fnl}{F_\mathrm{NL,iso}}

\newcommand{\abs}[1]{{\left \vert #1 \right \vert}}

\newcommand{\cA}{\mathcal{A}}

\newcommand{\cH}{\mathcal{H}}
\newcommand{\cS}{\mathcal{S}}
\newcommand{\cO}{\mathcal{O}}

\newcommand{\bk}{\mathbf{k}}

\newcommand{\bq}{\mathbf{q}}

\newcommand{\bx}{\mathbf{x}}

\newcommand{\ud}{\mathrm{d}}

\newcommand{\uGW}{\mathrm{gw}}

\newcommand{\Beq}{\begin{align}}
\newcommand{\Eeq}{\end{align}}

\DeclareAcronym{GW}{
  short = GW,
  long = gravitational wave ,
  short-plural = s ,
}
\DeclareAcronym{LIGO}{
  short =LIGO ,
  long = Laser Interferometer Gravitational-Wave Observatory ,
  short-plural = ,
}
\DeclareAcronym{LVK}{
  short = LVK ,
  long = {LIGO, Virgo, and KAGRA},
  short-plural = ,
}

\DeclareAcronym{SGWB}{
  short = SGWB ,
  long = stochastic gravitational-wave background ,
  short-plural = s ,
}
\DeclareAcronym{GWB}{
  short = GWB ,
  long = gravitational-wave background ,
  short-plural = s ,
}
\DeclareAcronym{CBC}{
  short = CBC ,
  long = compact binary coalescence ,
  short-plural = s ,
}
\DeclareAcronym{BH}{
  short = BH ,
  long = black hole ,
  short-plural = s ,
}
\DeclareAcronym{BBH}{
  short = BBH ,
  long = binary black hole ,
  short-plural = s ,
}
\DeclareAcronym{PBH}{
  short = PBH ,
  long = primordial black hole ,
  short-plural = s ,
}
\DeclareAcronym{SNR}{
  short = SNR ,
  long = signal-to-noise ratio ,
  short-plural = s ,
}
\DeclareAcronym{IMRPPv2}{
  short = ,
  long = {\normalsize IMRP}{\footnotesize HENOM}{\normalsize P}v2 ,
  short-plural = ,
}
\DeclareAcronym{PTA}{
  short = PTA ,
  long = pulsar timing array ,
  short-plural = s ,
}
\DeclareAcronym{SFR}{
  short = SFR ,
  long = star formation rate ,
  short-plural =  ,
}
\DeclareAcronym{FRW}{
  short = FRW ,
  long = Friedmann-Robertson-Walker ,
  short-plural =  ,
}
\DeclareAcronym{IMR}{
  short = IMR ,
  long = inspiral-merger-ringdown ,
  short-plural =  ,
}
\DeclareAcronym{LISA}{
	short = LISA ,
	long  = Laser Interferometer Space Antenna,
  short-plural =  ,
}
\DeclareAcronym{ET}{
	short = ET ,
	long  = Einstein Telescope,
  short-plural =  ,
}
\DeclareAcronym{CE}{
	short = CE ,
	long  = Cosmic Explorer,
  short-plural =  ,
}
\DeclareAcronym{BBO}{
	short = BBO ,
	long  = Big Bang Observer,
  short-plural =  ,
}
\DeclareAcronym{DECIGO}{
	short = DECIGO ,
	long  = Deci-hertz Interferometer Gravitational wave Observatory,
  short-plural =  ,
}
\DeclareAcronym{ABH}{
	short = ABH ,
	long  = astrophysical black hole,
  short-plural = s ,
}

\DeclareAcronym{PNG}{
	short = PNG ,
	long  = primordial non-Gaussianity ,
  short-plural =  ,
}

\DeclareAcronym{CMB}{
	short = CMB ,
	long  = cosmic microwave background ,
  short-plural =  ,
}
\DeclareAcronym{LSS}{
	short = LSS ,
	long  = large-scale structure ,
  short-plural =  ,
}

\DeclareAcronym{PGW}{
	short = PGW ,
	long  = primordial gravitational wave ,
  short-plural = s ,
}
\DeclareAcronym{SIGW}{
	short = SIGW ,
	long  = scalar-induced gravitational wave ,
  short-plural = s ,
}

\DeclareAcronym{RD}{
	short = RD,
	long  = radiation-dominated ,
  short-plural =  ,
}
\DeclareAcronym{eRD}{
	short = eRD,
	long  = early radiation-dominated ,
  short-plural =  ,
}
\DeclareAcronym{MD}{
	short = MD,
	long  = matter-dominated ,
  short-plural =  ,
}
\DeclareAcronym{eMD}{
	short = eMD,
	long  = early matter-dominated ,
  short-plural =  ,
}
\DeclareAcronym{SW}{
	short = SW,
	long  = Sachs-Wolfe ,
  short-plural =  ,
}
\DeclareAcronym{ISW}{
	short = ISW,
	long  = integrated Sachs-Wolfe ,
  short-plural =  ,
}

\DeclareAcronym{DM}{
	short = DM,
	long  = dark matter ,
  short-plural =  ,
}

\DeclareAcronym{NANOGrav}{
	short = NANOGrav ,
	long  = North American Nanohertz Observatory for Gravitational Waves ,
  short-plural =  ,
}

\DeclareAcronym{PDF}{
	short = PDF ,
	long  = probability distribution function ,
  short-plural = s ,
}

\DeclareAcronym{SMBH}{
  short = SMBH ,
  long  = supper-massive black hole ,
  short-plural = s ,
}
\DeclareAcronym{SKA}{
	short = SKA ,
	long  = Square Kilometre Array,
  short-plural =  ,
}
\DeclareAcronym{NG15}{
  short = NG15 ,
  long  = NANOGrav 15-year ,
  short-plural =  ,
}

\title{\boldmath Isotropic Background and Anisotropies of Gravitational Waves Induced by Cosmological Soliton Isocurvature Perturbations}

\author[a,b]{Di Luo,}
\emailAdd{luo1844725234@mail.ustc.edu.cn}
\author[a,c,\ast]{Yan-Heng Yu,} 
\emailAdd{yhyu@ihep.ac.cn}
\author[a,c,\ast]{Jun-Peng Li,}
\emailAdd{lijunpeng@ihep.ac.cn}
\author[d,a,c,\ast]{Sai Wang\note[$\ast$]{Corresponding author.}} 
\emailAdd{wangsai@ihep.ac.cn}

\affiliation[a]{Theoretical Physics Division, Institute of High Energy Physics, Chinese Academy of Sciences, 19B Yuquan Road, Shijingshan District, Beijing 100049, China}
\affiliation[b]{School of Physical Sciences, University of Science and Technology of China, 96 Jinzhai Road, Baohe District, Hefei 230026, China}
\affiliation[c]{School of Physics, University of Chinese Academy of Sciences, 19A Yuquan Road, Shijingshan District, Beijing 100049, China}
\affiliation[d]{Department of Physics, Hangzhou Normal University, Hangzhou 310036, P.R.China}

\abstract{
Cosmological solitons are widely predicted by scenarios of the early Universe. In this work, we investigate the isotropic background and anisotropies of gravitational waves (GWs) induced by soliton isocurvature perturbations, especially considering the effects of non-Gaussianity in these perturbations. Regardless of non-Gaussianity, the energy-density fraction spectrum of isocurvature-induced GWs approximately has a universal shape within the perturbative regime, thus serving as a distinctive signal of solitons. We derive the angular power spectrum of isocurvature-induced GWs to characterize their anisotropies. Non-Gaussianity plays a key role in generating anisotropies through the couplings between large- and small-scale isocurvature perturbations, making the angular power spectrum to be a powerful probe of non-Gaussianity. Moreover, the isocurvature-induced GWs have nearly no cross-correlations with the cosmic microwave background, providing a new observable to distinguish them from other GW sources, e.g., GWs induced by cosmological curvature perturbations enhanced at small scales. Therefore, detection of both the isotropic background and anisotropies of isocurvature-induced GWs could reveal important implications for the solitons as well as the early Universe.
}

\begin{document}

\maketitle
\flushbottom
\allowdisplaybreaks

\section{Introduction}

The long-lived localized massive objects, known as ``solitons'', are prevalent in a variety of scenarios of the early Universe and have garnered significant interest.
These solitons, including topological defects, oscillons, Q-balls, etc., can naturally arise in cosmological models containing new physics (e.g., see review \cite{Zhou:2024mea} and references therein), or are introduced to address fundamental issues like dark matter (e.g., see Ref.~\cite{Murayama:2009nj}).
Exploring the properties of solitons holds essential implications for comprehending the early Universe.

The cosmological isocurvature perturbations contain rich information about solitons in the early Universe.
At the formation time, solitons generally have little impact on the total energy density of the Universe, primarily introducing relative inhomogeneities among different components, thereby acting as isocurvature perturbations.
These perturbations are orthogonal to the cosmological curvature perturbations, with the latter arising from the perturbations to the radiation that dominates the evolution of the early Universe.
At large scales, the spectral amplitude of isocurvature perturbations is less than $\sim 1\%$ of that of curvature ones, as constrained by observations of the \ac{CMB} \cite{Planck:2018jri}.
However, the universal characteristics of the solitons can lead to enhanced isocurvature perturbations at much smaller scales associated with their typical non-linear scales \cite{Lozanov:2023aez}, providing a valuable avenue for studying solitons.

Reflecting deviations from the Gaussian statistics, the non-Gaussianity is one of the most important natures of the soliton isocurvature perturbations.
Just as the non-Gaussianity in curvature perturbations provides insights into inflationary dynamics \cite{Maldacena:2002vr,Bartolo:2004if,Allen:1987vq,Bartolo:2001cw,Acquaviva:2002ud,Bernardeau:2002jy,Chen:2006nt,Gao:2012ib,Gao:2008dt,Huang:2008zj,Huang:2009vk,Cai:2018dkf}, the non-Gaussianity in soliton isocurvature perturbations offers a distinctive window into the underlying dynamics of the fields related to these solitons.
Detecting such non-Gaussianity at large scales would confront challenges since the isocurvature perturbations are constrained to be small.
On the other hand, detecting it at small scales, though lacking tight constraints, necessitates the development of new detection approaches.

\Acp{GW} induced by isocurvature perturbations provide a new probe of the solitons. 
In contrast to \acp{GW} induced by curvature perturbations, which are known as ``\acp{SIGW}'' \cite{Ananda:2006af,Baumann:2007zm,Mollerach:2003nq,Assadullahi:2009jc,Domenech:2021ztg,Espinosa:2018eve,Kohri:2018awv}, \acp{GW} can not be induced directly by isocurvature perturbations, since they do not affect the shape of space by definition.
However, if the energy-density fractions of the matter ingredients in the Universe undergo changes, the initial isocurvature perturbations may partially evolve into curvature perturbations, leading to the inevitable generation of \acp{GW} due to nonlinear effects of gravity \cite{Domenech:2021and}.
These so-called ``isocurvature-induced \acp{GW}'' can be generally produced from cosmological solitons \cite{Lozanov:2023aez}.
Since the solitons behave as the non-relativistic matter, their energy density decays slower than that of radiation as the Universe expands.
With the energy-density fraction of solitons growing, the initial soliton isocurvature perturbations at small scales transform into curvature ones, and act as the sources of \acp{GW}.
Other relevant works of the isocurvature-induced \acp{GW} can also be found in the literature \cite{Papanikolaou:2020qtd,Domenech:2020ssp,Lozanov:2023rcd,Domenech:2023jve,Lozanov:2023knf,Domenech:2024cjn,Domenech:2024kmh,Chen:2024twp,Domenech:2024wao,Yuan:2024qfz,Kumar:2024hsi,Papanikolaou:2024kjb,Dalianis:2021dbs,He:2024luf,Han:2025wlo}.

Identifying the non-Gaussianity in soliton isocurvature perturbations via the isocurvature-induced \acp{GW} remains one of the important research objectives.
This study will delve into both the isotropic background and anisotropies of isocurvature-induced \acp{GW}, with a specific focus on the effects of non-Gaussianity.
As for the isotropic background, Ref.~\cite{Lozanov:2023aez} shows that for a wide class of solitons, the corresponding isocurvature-induced \acp{GW} share the same shape of energy-density fraction spectrum, known as ``universal \acp{GW}''.
We will reveal that this energy-density fraction spectrum is insensitive to non-Gaussianity in isocurvature perturbations within the perturbative regime, indicating that this ``universal \acp{GW}'' could still serve as a distinctive signal for detecting these solitons, irrespective of whether the isocurvature perturbations are Gaussian or not.
This differs from the case of \acp{SIGW}, whose energy-density fraction spectrum can experience significant modifications due to the non-Gaussianity in primordial curvature perturbations \cite{Adshead:2021hnm,Ragavendra:2021qdu,Abe:2022xur,Yuan:2023ofl,Perna:2024ehx,Li:2023qua,Li:2023xtl,Ruiz:2024weh,Wang:2023ost,Yu:2023jrs,Iovino:2024sgs,Zeng:2024ovg,Cai:2018dig,Unal:2018yaa,Atal:2021jyo,Yuan:2020iwf,Chang:2023aba,Zhou:2024yke,Nakama:2016gzw,Garcia-Bellido:2017aan,Ragavendra:2020sop,Zhang:2021rqs,Lin:2021vwc,Chen:2022dqr,Cai:2019elf}.
To extract the information of non-Gaussianity, we will further study the anisotropies of isocurvature-induced \acp{GW} and derive the (reduced) angular power spectrum.
Analogous to the analysis of anisotropies in \acp{SIGW} \cite{Bartolo:2019zvb,Li:2023qua,Li:2023xtl,Rey:2024giu,Ruiz:2024weh,Schulze:2023ich,LISACosmologyWorkingGroup:2022kbp,LISACosmologyWorkingGroup:2022jok,Malhotra:2022ply,Wang:2023ost,Yu:2023jrs}, the non-Gaussianity in isocurvature perturbations will be shown to play a critical role in generating the anisotropies of isocurvature-induced \acp{GW}, and notably affect the angular power spectrum.
Therefore, these anisotropies could be an effective probe of the non-Gaussianity in isocurvature perturbations, breaking its degeneracy with other model parameters and shedding light on the dynamics associated with solitons.
Furthermore, we will demonstrate that  cross-correlations between \acp{GW} and the \ac{CMB}, could offer features distinct from other \ac{GW} sources, e.g., \acp{SIGW} \cite{Dimastrogiovanni:2022eir,Schulze:2023ich,Zhao:2024gan}, serving as novel signals to identify solitons.

The rest of this paper is structured as follows.
In \cref{sec:evo}, we review the cosmological isocurvature perturbations due to solitons.
In \cref{sec:ogw}, we investigate the energy-density fraction spectrum of isocurvature-induced \acp{GW}, considering the effects of non-Gaussianity in soliton isocurvature perturbations.
In \cref{sec:aps}, we analyze the angular power spectrum of isocurvature-induced \acp{GW}, and focus on its implications for detecting the non-Gaussianity in soliton isocurvature perturbations.
Moreover, we compare our results to those of \acp{SIGW}.
In \cref{sec:conc}, we reveal the conclusions and discussion.
In \cref{sec:inte}, we summarize the formulas for the energy-density fraction spectrum necessary for \cref{sec:ogw}.

\section{Cosmological isocurvature perturbations from solitons}
\label{sec:evo}

Similar to Ref.~\cite{Lozanov:2023aez}, we consider an early Universe comprising two fluids, namely, radiation originating from the inflaton decay, and solitons formed by a spectator field which is only weakly coupled to the inflaton.
When solitons form, the energy density of radiation greatly exceeds that of solitons, resulting in an \ac{RD} era.
While fluctuations in radiation contribute to cosmological curvature perturbations, the Poisson distribution of solitons gives rise to enhanced cosmological isocurvature perturbations at small scales.
As the energy-density fraction of solitons increases with the Universe's expansion, these isocurvature perturbations gradually evolve into curvature ones, acting as sources of \acp{GW} in the \ac{RD} era, termed ``isocurvature-induced \acp{GW}''.
Further, if solitons are long-lived enough, they could initiate a soliton-dominated era at some time. 
This soliton-dominated era would transit into the standard \ac{RD} era when solitons eventually decay, which may lead to significant production of \acp{GW}, depending on whether this transition is sudden or not \cite{Inomata:2019ivs,Inomata:2019zqy,Pearce:2023kxp}.
However, to avoid getting stuck in a discussion about the details of the soliton-dominated era, which is not the primary focus of our present paper, we assume that the solitons decay at around the radiation-soliton equality, resulting in the absence of soliton domination.

Let us focus on the cosmological perturbations first.
We consider a perturbed \acp{FRW} metric in the conformal Newtonian gauge, i.e.,
\begin{equation}\label{eq:metric}
    \ud s^2
    = a^2(\eta) \left\{
            - \left( 1 + 2 \Phi \right) \ud \eta^2
            + \left[ (1 - 2 \Phi) \delta_{ij} + \frac{1}{2} h_{ij} \right]
            \ud x^i \ud x^j
        \right\}\ .
\end{equation}
Here, $a(\eta)$ is a scale factor of the Universe at the conformal time $\eta$, and is given by $a(\eta)/a_\mathrm{eeq}=2\eta/\eta_\ast+(\eta/\eta_\ast)^2$ in the radiation-soliton Universe, where $\eta_\ast=(\sqrt{2}+1)\,\eta_\mathrm{eeq}$, and the subscript ``eeq'' denotes the radiation-soliton equality.
The $\Phi$ denotes the first-order cosmological scalar perturbations with the anisotropic stress being neglected, and is proportional to the linear cosmological curvature perturbations $\zeta$.
The $h_{ij}$ stands for the second-order transverse-traceless tensor perturbations, which will be studied in the subsequent sections.
Moreover, the cosmological isocurvature perturbations $\cS$ do not affect the spacetime metric, and is defined in a gauge-invariant form as \cite{Kodama:1984ziu,Malik:2008im}
\begin{equation}\label{eq:definitionS}
	\cS = \frac{\delta\rho_m}{\rho_m}-\frac{3}{4}\frac{\delta\rho_r}{\rho_r}\ ,
\end{equation}
where $\rho_m$ (or $\rho_r$) is the energy density of solitons (or radiation), and $\delta\rho_m$ (or $\delta\rho_r$) is the fluctuations in $\rho_m$ (or $\rho_r$). 
It should be noted that $\cS$ can be roughly determined by the density contrast of solitons during the \ac{RD} era, namely $\cS\simeq \delta\rho_m/\rho_m$, considering $\rho_m\ll\rho_r$ and the isocurvature condition $\delta\rho_m+\delta\rho_r=0$ in Eq.~(\ref{eq:definitionS}).

We investigate $\zeta$ and $\cS$ at the initial conformal time $\eta_\mathrm{i}$ when solitons form.
Different from Ref.~\cite{Lozanov:2023aez}, we consider the non-Gaussianity in $\cS$, which can result from some dynamic mechanism of the field related to soliton formation, whereas we neglect the non-Gaussianity in $\zeta$. 
We further assume that the non-Gaussian $\cS$ is parameterized in the Fourier space as
\footnote{
We consider the local-type non-Gaussianity in $\cS$ to establish our framework.
This type of non-Gaussianity can arise from certain soliton scenarios, e.g., those involving solitons with conserved charges.
Assuming the linear energy-charge relation for solitons, the energy density of solitons at a coarse-grained location approximately scale proportionally with the charge number density.
If the charge number perturbations exhibit local-type non-Gaussian statistics due to the underlying non-linear dynamics of some field, the resulting soliton isocurvature perturbations could follow the same local-type non-Gaussian statistics.
A specific example can be thin-wall Q-balls carrying baryon number, whose perturbations could display local-type non-Gaussian statistics through Affleck-Dine baryogenesis \cite{Kawasaki:2008jy}.
}
\begin{equation}\label{eq:fnl-def-k}
    \cS (\eta_\mathrm{i},\bk) = \cS_g(\eta_\mathrm{i},\bk) +  \Fnl  \int \frac{\ud^3 \bq}{(2\pi)^{3/2}} \cS_g(\eta_\mathrm{i},\bq) \cS_g(\eta_\mathrm{i},\bk-\bq)\ ,
\end{equation}
where $\bk$ and $\bq$ are the Fourier modes of perturbations, $\cS_{g}$ stands for the Gaussian component of $\cS$, and $\Fnl$ is the non-linearity parameter of $\cS$. 
In this work, we focus solely on the non-Gaussianity up to the order of $\Fnl$, but the effects of higher-order non-Gaussianity can be straightforwardly extended to, following the framework established in Refs.~\cite{Abe:2022xur,Li:2023xtl,Yuan:2023ofl,Perna:2024ehx,Ruiz:2024weh}.
In order to describe the statistics of Gaussian variables $\zeta$ and $\cS_g$, we introduce their dimensionless power spectra as follows
\begin{eqnarray}
   &&\langle \zeta (\eta_\mathrm{i},\bk_1) \cS_{g} (\eta_\mathrm{i},\bk_2) \rangle
    = 0\label{eq:PzetaS}\ ,\\
    &&\langle \zeta (\eta_\mathrm{i},\bk_1) \zeta (\eta_\mathrm{i},\bk_2) \rangle
    = \delta^{(3)} (\bk_1+\bk_2) \frac{2 \pi^2}{k_1^3}
    \Delta^2_{\zeta} (k_1)\ ,\label{eq:Pzeta} \\
    &&\langle \cS_{g} (\eta_\mathrm{i},\bk_1) \cS_{g} (\eta_\mathrm{i},\bk_2) \rangle
    = \delta^{(3)} (\bk_1+\bk_2) \frac{2 \pi^2}{k_1^3}
    \Delta^2_{\cS_g} (k_1)\ ,\label{eq:PS}
\end{eqnarray}
where we have $k=|\bk|$, and $\langle...\rangle$ represents the ensemble average.
In this work, the correlator in Eq.~(\ref{eq:PzetaS}) is expected to vanish, since $\zeta$ and $\cS_g$ are assumed to have uncorrelated origins.
We adopt a scale-invariant $\Delta^2_{\zeta}$ in Eq.~(\ref{eq:Pzeta}) with the spectral amplitude $\cA_{\mathrm{ad}}\simeq 2.1\times 10^{-9}$ that is revealed by the Planck 2018 results \cite{Planck:2018vyg}.
In contrast, $\Delta^2_{\cS_g}$ in Eq.~(\ref{eq:PS}) is modeled as
\begin{equation}\label{eq:power-law}
    \Delta^2_{\cS_g} (k) 
    = 
    \cA_{L,\mathrm{iso}}
    +
    \cA_{S,\mathrm{iso}}
    \left(\frac{k}{k_{\mathrm{uv}}}\right)^3 \Theta (k_{\mathrm{uv}}-k)\ ,
\end{equation}
where the subscript ``$L$'' (or ``$S$'') represents large (or small) scales, and a cutoff scale $k_\mathrm{uv}$ is introduced to maintain the validity of our ``soliton fluid" description. 
At large scales, $\cS_{gL}$ has a scale-invariant spectral amplitude $\cA_{L,\mathrm{iso}}$ with the constraint $\beta=\cA_{L,\mathrm{iso}}/\cA_{\mathrm{ad}} \lesssim 10^{-2}$ based on observations of the \ac{CMB} temperature anisotropies and polarization
\footnote{
As a concrete example, the authors of Ref.~\cite{Lozanov:2023aez} explore a model involving oscillons from an axion-like particle field and consider a specific parameter choice to produce nearly scale-invariant large-scale isocurvature perturbations with the spectral amplitude saturating the \ac{CMB} bound.
}
\cite{Planck:2018jri}.
At small scales $k\lesssim k_\mathrm{uv}$, $\cS_{gS}$ follows a $k^3$-spectrum due to the Poisson distribution of solitons, and the spectral amplitude at $k=k_{\mathrm{uv}}$ can be enhanced to $\cA_{S,\mathrm{iso}} \sim \cO(1)$ \cite{Lozanov:2023aez}.

We then focus on how the small-scale isocurvature perturbations evolve into curvature ones at $\eta$, which satisfies $\eta_\mathrm{i}<\eta \ll\eta_\mathrm{eeq}$.
Deep in the \ac{RD} era, the equations of motion of $\Phi$ and $\cS$ are given by \cite{Domenech:2021and,Lozanov:2023aez}
\begin{eqnarray}
	\frac{\ud^2\Phi}{\ud x^2}+\frac{4}{x}\frac{\ud\Phi}{\ud x}+\frac{1}{3}\Phi
	+\frac{1}{4\sqrt{2}\kappa x} \left[x\frac{\ud\Phi}{\ud x}+(1-x^2)\Phi-2\cS\right]
	&&\simeq 0\ ,\label{eq:evoPhi}\\
	\frac{\ud^2\cS}{\ud x^2}+\frac{1}{x}\frac{\ud \cS}{\ud x}-\frac{x^2}{6}\Phi
	-\frac{1}{2\sqrt{2}\kappa}\left(\frac{\ud\cS}{\ud x}-\frac{x}{2}\cS-\frac{x^3}{12}\Phi\right)
	&&\simeq 0\ .\label{eq:evoS}
\end{eqnarray}
Here, we introduce the dimensionless parameters $x=k\eta\gg1$ and $\kappa=k/k_{\rm eeq}$ with $x/\kappa\ll 1$, where $k_\mathrm{eeq}$ is related to the mode reentering the horizon at $\eta_\mathrm{eeq}$.
To solve the above equations, we approximately set the initial conditions as $\Phi=0$ and $\cS=\cS(\eta_\mathrm{i},\bk)$ because of $\cA_{\mathrm{ad}}\ll\cA_{S,\mathrm{iso}}$.
The analytical solutions of Eqs.~\eqref{eq:evoPhi} and \eqref{eq:evoS} up to order $(x/\kappa)^{2}$ are given by \cite{Domenech:2021and,Lozanov:2023aez}
\begin{align}
	\Phi(\eta,\bk) &\simeq
	\frac{3\,\cS(\eta_\mathrm{i},\bk)}{2\sqrt{2}\kappa} \frac{1}{x^3}
	\left[6+x^2-2\sqrt{3}x\sin\left(\frac{x}{\sqrt{3}}\right)-6\cos\left(\frac{x}{\sqrt{3}}\right)\right]\ ,\label{eq:solPhi}\\
	\cS (\eta,\bk) &\simeq
	\cS(\eta_\mathrm{i},\bk)+\frac{3\,\cS(\eta_\mathrm{i},\bk)}{2\sqrt{2}\kappa}
	\left[x+\sqrt{3}\sin\left(\frac{x}{\sqrt{3}}\right)-2\sqrt{3}\, {\rm Si}\left(\frac{x}{\sqrt{3}}\right)\right]\ ,\label{eq:solS}
\end{align}
where ${\rm Si}(x)$ is the sine-integral function.
Though these solutions are derived in the condition $\eta_\mathrm{i}=0$, they are still applicable for the case $\eta_\mathrm{i}>0$ \cite{Lozanov:2023aez}.
As demonstrated in Eq.~(\ref{eq:solPhi}), the $\cS(\eta_\mathrm{i},\bk)$ could evolve into the scalar perturbations $\Phi$ or correspondingly the curvature perturbations $\zeta$, which are the sources of \acp{GW} studied below.

\section{Isotropic background of isocurvature-induced GWs}
\label{sec:ogw}

As the soliton isocurvature perturbations evolve into curvature perturbations during the \ac{RD} era, the subsequent evolution of the latter inside the horizon inevitably induces \acp{GW}, known as "isocurvature-induced \acp{GW}", i.e., the $h_{ij}$ in Eq.~(\ref{eq:metric}).
The energy density of these \acp{GW} is defined by
\begin{equation}\label{eq:rhogw}
    \rho_\mathrm{gw}(\eta,\bx)=\frac{M_\mathrm{pl}^2}{16\,a^2(\eta)}\,
    \overbar{\partial_l h_{ij}(\eta,\bx)\, \partial_l h_{ij}(\eta,\bx) }\ ,
\end{equation}
where $\bx$ is a spatial position, the long overbar represents oscillation average, and $M_\mathrm{pl}$ is the reduced Planck mass.
We further decompose $\rho_\mathrm{gw}$ as
\begin{equation}
    \rho_\mathrm{gw}(\eta,\bx)=\bar{\rho}_\mathrm{gw}(\eta)+\delta\rho_\mathrm{gw}(\eta,\bx)\ ,
\end{equation} 
where $\bar{\rho}_\mathrm{gw}$ refers to the isotropic background, i.e., the spatial average of $\rho_\mathrm{gw}(\eta,\bx)$, and $\delta\rho_\mathrm{gw}$ denotes the fluctuations on top of this background.
In order to characterize $\bar{\rho}_\mathrm{gw}$ and $\delta\rho_\mathrm{gw}$, we define the energy-density fraction spectrum $\bar{\Omega}_\uGW (\eta,\bx)$ and the density contrast $\delta_\uGW (\eta,\bx,\bk)$, respectively, i.e., \cite{Bartolo:2019oiq,Bartolo:2019yeu}
\begin{subequations}
\begin{align}
    \bar{\Omega}_\uGW (\eta,k) 
    &= \frac{1}{\rho_c(\eta)}\,
    \frac{\ud\, \bar{\rho}_\uGW (\eta)}{ \ud\, \ln{k}}\ ,\label{eq:Omega}
    \\
    \,\delta_\uGW (\eta,\bx,\bk)
    &=
    \frac{1}{\rho_c(\eta)} \frac{4\pi}{\bar{\Omega}_\mathrm{gw}(\eta,k)}\,
    \frac{\ud\, \delta\rho_\uGW (\eta,\bx) }{ \ud\, \ln{k} \  \ud^2 \,\hat{\bk}}\ , \label{eq:delta}
\end{align}    
\end{subequations}
where $\rho_c(\eta)$ is the critical energy density of the Universe at $\eta$, $\bk$ is the wavevector of \acp{GW} with $\hat{\bk}=\bk/k$ being the unit directional vector.

In this section, we focus on $\bar{\Omega}_\uGW $, which describes the homogeneous and isotropic background of \acp{GW}.
In the subsequent section, we will study $\delta_\uGW (\eta,\bx,\bk)$, which results in the inhomogeneities and anisotropies in \acp{GW}.

\subsection{Energy-density fraction spectrum}

The energy-density fraction spectrum of isocurvature-induced \acp{GW} can be computed using a semi-analytical approach akin to that employed for \acp{SIGW} \cite{Ananda:2006af,Baumann:2007zm,Mollerach:2003nq,Assadullahi:2009jc,Domenech:2021ztg,Espinosa:2018eve,Kohri:2018awv}. 
In Fourier space, $h_{ij}$ is expressed as
\begin{equation}
    h_{ij}(\eta,\bx)
    = \sum_{\lambda=+,\times}
        \int \frac{\ud^3 \bk}{(2\pi)^{3/2}} e^{i\bk\cdot\bx}
        \epsilon_{ij,\bk}^{\lambda}\, h_\lambda(\eta, \bk)\ ,
\end{equation}
where $\lambda=+,\times$ are two polarization modes of \acp{GW}, and the polarization tensors are defined as $\epsilon^+_{ij,\bk}=[\epsilon_{i,\bk} \epsilon_{j,\bk}- \bar{\epsilon}_{i,\bk} \bar{\epsilon}_{j,\bk}]/\sqrt{2}$ and $\epsilon^\times_{ij,\bk}=[\epsilon_{i,\bk} \epsilon_{j,\bk}+ \bar{\epsilon}_{i,\bk} \bar{\epsilon}_{j,\bk}]/\sqrt{2}$ with $\epsilon_{i,\bk}$ and $\bar{\epsilon}_{i,\bk}$ forming an orthonormal basis perpendicular to $\mathbf{k}$.
The equation of motion of $h_{\lambda}(\eta,\bk)$ is given by
\begin{equation}\label{eq:SIGW-motion}
    h''_{\lambda}(\eta,\bk)
    + 2\cH h'_{\lambda}(\eta,\bk)
    +k^2 h_{\lambda}(\eta,\bk) 
    = 4 \mathcal{T}_{\lambda}(\eta,\bk)\ ,
\end{equation} 
where a prime stands for a derivative with respect to $\eta$. 
The source term $\mathcal{T}_{\lambda}(\eta,\bk)$ is given by
\begin{eqnarray}\label{eq:Slam}
    \mathcal{T}_\lambda(\eta, \mathbf{k})
    = \int \frac{\ud^3 \bq}{(2\pi)^{3/2}} Q_{\lambda}(\mathbf{k},\mathbf{q})
        f(\vert \mathbf{k}-\mathbf{q} \vert, q, \eta)
        \cS(\eta_\mathrm{i},\bk-\bq) \cS(\eta_\mathrm{i},\bq)\ ,
\end{eqnarray}
where $Q_{\lambda}(\mathbf{k}, \mathbf{q}) = \epsilon_{ij}^{\lambda}(\mathbf{k})\, q_i q_j$ denotes a projection factor, and $f(\vert \bk-\bq \vert, q, \eta)$ is given by
\begin{equation}\label{eqn:source-function}
    f(p, q, \eta)
    = 2 T(p,\eta) T(q,\eta) +
        \left[T(p,\eta) +\eta T'(p,\eta)\right]
        \left[T(q,\eta) +\eta T'(q,\eta)\right]\ ,
\end{equation}
where we introduce $p=\vert \mathbf{k}-\mathbf{q} \vert$, and the transfer function $T(\eta,k)$ connects $\Phi(\eta,\bk)$ and the initial isocurvature perturbations $\cS(\eta_\mathrm{i},\bk)$, i.e.,
\begin{equation}\label{eq:T-def}
    \Phi(\eta, \bk)
        = T(\eta,k) \cS(\eta_\mathrm{i},\bk)\ .
\end{equation}
The explicit expression of $T(\eta,k)$ can be straightforwardly read from Eq.~\eqref{eq:solPhi}, i.e.,
\begin{equation}\label{eq:T-iso}
    T(\eta,k) \simeq \frac{3}{2\sqrt{2}x^3\kappa} \left[6+x^2-2\sqrt{3}x\sin\left(\frac{x}{\sqrt{3}}\right)-6\cos\left(\frac{x}{\sqrt{3}}\right)\right]\ .
\end{equation}
Following the Green's function method \cite{Espinosa:2018eve,Kohri:2018awv}, the solution to \cref{eq:SIGW-motion} is formally given by   
\begin{equation}\label{eq:h}
    h_\lambda(\eta, \bk)
   =\frac{4}{k^2} \int \frac{\ud^3 \bq}{(2\pi)^{3/2}}\ 
   Q_{\lambda}(\bk,\bq)\,
   \hat{I} \left(\frac{\abs{\bk-\bq}}{k},\frac{q}{k},\kappa,x\right)
   \cS(\eta_\mathrm{i},{\bk-\bq})\cS(\eta_\mathrm{i},{\bq})\ ,
\end{equation}
where the kernel function $\hat{I}(u,v,\kappa,x)$ with $x \gg 1$ is given by
\footnote{
The analytical expressions in Eq.~(\ref{eq:I-RD}) are derived under the condition $k_{\mathrm{uv}} \eta_{\mathrm{i}}\rightarrow 0$.
For $k_{\mathrm{uv}} \eta_{\mathrm{i}}\gtrsim 1$, the kernel function lacks simple expressions.
In this work, we adopt the result in Eq.~(\ref{eq:I-RD}) as an approximation.
Numerical analysis indicates that the change of the energy-density fraction spectrum within the range of $0<k_{\mathrm{uv}} \eta_{\mathrm{i}} \lesssim10$ is less than $\sim 30\%$, which would not affect our main results.
}
\cite{Domenech:2021and}
\begin{subequations}\label{eq:I-RD}
\begin{eqnarray}
    \hat{I} (u,v,\kappa,x \gg 1) &=& \frac{1}{x} I_A (u,v,\kappa) 
        \left[I_B (u,v) \sin x - \pi I_C (u,v) \cos x\right]\ ,\\
    I_A (u,v,\kappa) &=& 9/\left(16 u^4 v^4 \kappa^2\right)\ , \\
    I_B (u,v) &=& -3 u^2 v^2+\left(-3 + u^2\right)
        \left(-3 + u^2 + 2v^2\right)
        \ln\left|1-\frac{u^2}{{3}}\right|\nonumber\\
        & &\ +\left(-3+v^2\right)\left(-3+v^2+2u^2\right)\ln\left|1-\frac{v^2}{3}\right|\nonumber\\
        & &\ -\frac{1}{2}\left(-3+v^2+u^2\right)^2
        \ln\left[\,\left|1-\frac{(u+v)^2}{{3}}\right|\left|1-\frac{(u-v)^2}{3}\right|\,\right]\ , \\
    I_C (u,v) &=& 9-6v^2-6u^2+2u^2v^2\nonumber\\
        & &\ +\left(3-u^2\right)\left(-3+u^2+2v^2\right)\Theta\left(1-\frac{u}{\sqrt{3}}\right)\nonumber\\
        & &\ +\left(3-v^2\right)\left(-3+v^2+2u^2\right)\Theta\left(1-\frac{v}{\sqrt{3}}\right)\nonumber\\
        & &\ +\frac{1}{2}\left(-3+v^2+u^2\right)^2
        \left[\Theta\left(1-\frac{u+v}{\sqrt{3}}\right)+\Theta\left(1+\frac{u-v}{\sqrt{3}}\right)\right]\ .
\end{eqnarray}
\end{subequations}
Note that the kernel function of isocurvature-induced \acp{GW} exhibits a factor of $\kappa^{-2}$, which is different from the case for \acp{SIGW} \cite{Espinosa:2018eve,Kohri:2018awv}.

At the conformal time of \ac{GW} emission, denoted as $\eta_\mathrm{e}$, the energy-density fraction spectrum of isocurvature-induced \acp{GW}, denoted as $\bar{\Omega}_\mathrm{gw,e}(k) = \bar{\Omega}_\mathrm{gw}(\eta_\mathrm{e}, k)$ for simplicity, can be obtained from Eq.~(\ref{eq:Omega}), i.e.,
\begin{equation}\label{eq:energy-density}
    \bar{\Omega}_\mathrm{gw,e}(k)
    = \frac{1}{48}
        \left[\frac{k}{\cH(\eta_\mathrm{e})}\right]^2
        \sum_{\lambda=+,\times} \overbar{\Delta^2_\lambda(\eta_\mathrm{e},k)} \ ,
\end{equation}
where $\cH(\eta_\mathrm{e})$ is the conformal Hubble parameter at $\eta_\mathrm{e}$, and the dimensionless power spectrum of \acp{GW} is defined as
\begin{equation}\label{eq:Deltah}
    \langle
        h_{\lambda_1}(\eta_\mathrm{e}, \bk_1)
        h_{\lambda_2}(\eta_\mathrm{e}, \bk_2)
    \rangle
    = \delta_{\lambda_1 \lambda_2}\,\delta^{(3)}(\bk_1 + \bk_2)\,
        \frac{2\pi^2}{k_1^3}\, \Delta^2_{\lambda_1}(\eta_\mathrm{e}, k_1)\ .
\end{equation}
Based on Eqs.~(\ref{eq:h}), (\ref{eq:energy-density}), and (\ref{eq:Deltah}), we have $\bar{\Omega}_\mathrm{gw,e}\sim \langle h^2 \rangle \sim \langle \cS^4\rangle\simeq\langle \cS_S^4\rangle$, where the non-Gaussian $\cS_S$ is given by $\cS_S \sim \cS_{gS}+\Fnl\, \cS^2_{gS}$ based on Eq.~(\ref{eq:fnl-def-k}), and we can safely neglect the contribution from $\cS_{gL}$ because of $\cA_{L,\mathrm{iso}}\ll\cA_{S,\mathrm{iso}}$.
According to the Wick's theorem, the four-point correlator $\langle \cS_S^4\rangle$ can be expanded in terms of the two-point correlator $\langle \cS_{gS}^2\rangle$.
To simplify this calculation, we could apply a diagrammatic approach \cite{Adshead:2021hnm,Ragavendra:2021qdu,Abe:2022xur, Li:2023qua,Li:2023xtl,Li:2024zwx,Perna:2024ehx,Ruiz:2024weh,Rey:2024giu,Chang:2023aba,Chang:2023vjk,Zhou:2024ncc}, which is well-established in the study of the non-Gaussianity of curvature perturbations in the case of \acp{SIGW}.  
As a result, $\bar{\Omega}_\mathrm{gw}$ is expressed as
\begin{equation}\label{eq:Omegabar-total}
    \bar{\Omega}_\mathrm{gw,e}  
    = \bar{\Omega}_\mathrm{gw,e} ^{(0)}+\bar{\Omega}_\mathrm{gw,e} ^{(1)}+\bar{\Omega}_\mathrm{gw,e} ^{(2)} \ ,
\end{equation}
where $\bar{\Omega}_\mathrm{gw,e} ^{(n)}$ ($n=0,1,2$) denotes the  $\cO(\cA_{S,\mathrm{iso}}^{n+2} \Fnl^{2n})$ component of $\bar{\Omega}_\mathrm{gw,e}$, i.e., 
\begin{subequations}\label{eq:OmegaX}
\begin{align}
    \bar{\Omega}_\mathrm{gw,e} ^{(0)}
    &=\bar{\Omega}_\mathrm{gw,e} ^{G}\ ,
    \\
    \bar{\Omega}_\mathrm{gw,e} ^{(1)}
    &=\bar{\Omega}_\mathrm{gw,e} ^{H}+\bar{\Omega}_\mathrm{gw,e} ^{C}+\bar{\Omega}_\mathrm{gw,e} ^{Z}\ ,
    \\
    \bar{\Omega}_\mathrm{gw,e} ^{(2)}
    &=\bar{\Omega}_\mathrm{gw,e} ^{R}+\bar{\Omega}_\mathrm{gw,e} ^{P}+\bar{\Omega}_\mathrm{gw,e} ^{N}\ . 
\end{align}    
\end{subequations}
The explicit expressions of $\bar{\Omega}_\mathrm{gw,e}^X$ ($X=G,H,C,Z,R,P,N$) are listed in \cref{sec:inte}.
We note that their expressions share the same form as those of \acp{SIGW} except for the different specific expressions of the kernel functions \cite{Adshead:2021hnm,Ragavendra:2021qdu,Abe:2022xur,Li:2023qua,Li:2023xtl,Yuan:2023ofl,Perna:2024ehx,Ruiz:2024weh,Iovino:2024sgs,Wang:2023ost,Yu:2023jrs,Zeng:2024ovg,Garcia-Saenz:2022tzu}.

\subsection{Results}

\begin{figure*}[htbp]
    \centering
    \includegraphics[width=0.8\textwidth]{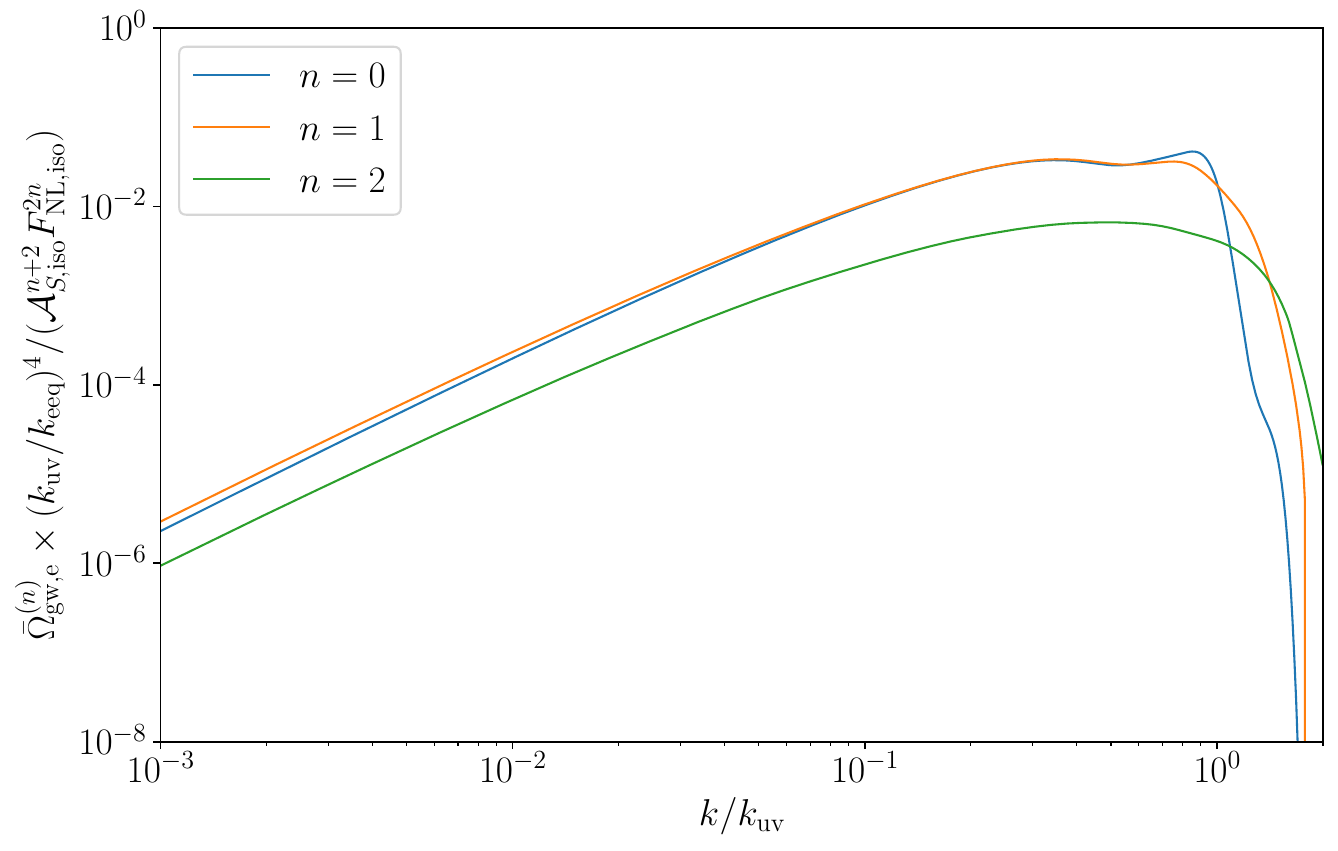}
    \caption{Unscaled components of the energy-density fraction spectrum of isocurvature-induced \acp{GW} as a function of the scale $k/k_\mathrm{uv}$.}
    \label{fig:Omega_order}
\end{figure*}

In Fig.~\ref{fig:Omega_order}, we present the unscaled components of the energy-density fraction spectrum of isocurvature-induced \acp{GW}, namely $\bar{\Omega}^{(n)}_\mathrm{gw,e}\times(k_\mathrm{uv}/k_\mathrm{eeq})^4/(\cA_{S,\mathrm{iso}}^{n+2} \Fnl^{2n})$ for $n=0,1,2$, as a function of the scale $k/k_{\mathrm{uv}}$.
As shown in \cref{fig:Omega_order}, each unscaled component shares a similar shape in the infrared regime $k\sim k_\mathrm{uv}$, and exhibits a rapid decline in the ultraviolet regime $k\gtrsim k_\mathrm{uv}$, which results from the unphysical cutoff $\Theta (k_{\mathrm{uv}}-k)$ in $\Delta^2_{\cS_g}$.
With the exception of the unphysical ultraviolet region, the unscaled components for $n=0$ and $n=1$ demonstrate comparable magnitudes, both surpassing the unscaled component for $n=2$.
It indicates that within the regime of $\cA_{S,\mathrm{iso}} \Fnl^2 <1$, the non-Gaussian contribution would enhance $\bar{\Omega}_\mathrm{gw,e}$ of isocurvature-induced \acp{GW}, but not significantly.
In particular, the non-Gaussian contributions are almost negligible for $\cA_{S,\mathrm{iso}} \Fnl^2 \lesssim\cO(0.1)$.
These results are different from those of \acp{SIGW}, where the non-Gaussianity of curvature perturbations could notably change the energy-density fraction spectrum of \acp{SIGW} \cite{Adshead:2021hnm,Ragavendra:2021qdu,Abe:2022xur,Yuan:2023ofl,Perna:2024ehx,Li:2023qua,Li:2023xtl,Ruiz:2024weh,Wang:2023ost,Yu:2023jrs,Iovino:2024sgs,Zeng:2024ovg,Cai:2018dig,Unal:2018yaa,Atal:2021jyo,Yuan:2020iwf,Chang:2023aba,Zhou:2024yke,Nakama:2016gzw,Garcia-Bellido:2017aan,Ragavendra:2020sop,Zhang:2021rqs,Lin:2021vwc,Chen:2022dqr,Cai:2019elf}.

The infrared behavior of the $\bar{\Omega}_\mathrm{gw,e}$ is a key feature of isocurvature-induced \acp{GW}. 
Following the method in Refs.~\cite{Yuan:2019wwo,Yuan:2023ofl,Cai:2018dig,Cai:2019cdl,Adshead:2021hnm}, we obtain that all the components exhibit the following behavior in the infrared regime $k\ll k_\mathrm{uv}$, i.e.,
\begin{equation}\label{eq:IR}
    \bar{\Omega}^{(n)}_\mathrm{gw,e}(k)\sim \left(\frac{k}{k_\mathrm{uv}}\right)^3\, \ln^2\left(\frac{k^2_\mathrm{uv}}{6 k^2}\right)\ .
\end{equation}
To derive the result, we note that the integrals in Eqs.~(\ref{eq:A1}-\ref{eq:A7}) are mainly contributed by the region $u_i, v_i \sim k_\mathrm{uv} / k$, and we also simplify these integrals by using the mean value theorem. 
Based on Eq.~(\ref{eq:IR}), the spectral index is given by $n_\mathrm{gw}=3-4/\ln{(k_\mathrm{uv}^2/6k^2)}$, which shows that both Gaussian and non-Gaussian components eventually decay as $k^3$ in the limit $k/k_\mathrm{uv}\rightarrow 0$.
For the Gaussian case, the infrared behavior of $\bar{\Omega}_\mathrm{gw,e}$ is also discussed in Ref.~\cite{Han:2025wlo}.
We find that even if non-Gaussianity is considered, the characteristic log-dependent infrared behavior remains unchanged, providing a helpful way to distinguish isocurvature-induced \acp{GW} from other \ac{GW} origins (e.g., for \acp{SIGW}, the spectral index in the infrared regime is given by $n_\mathrm{gw}=3-4/\ln{(4 k_\mathrm{uv}^2/3k^2)}$  \cite{Yuan:2019wwo,Yuan:2023ofl}).

\begin{figure*}[htbp]
    \centering
    \includegraphics[width=0.8\textwidth]{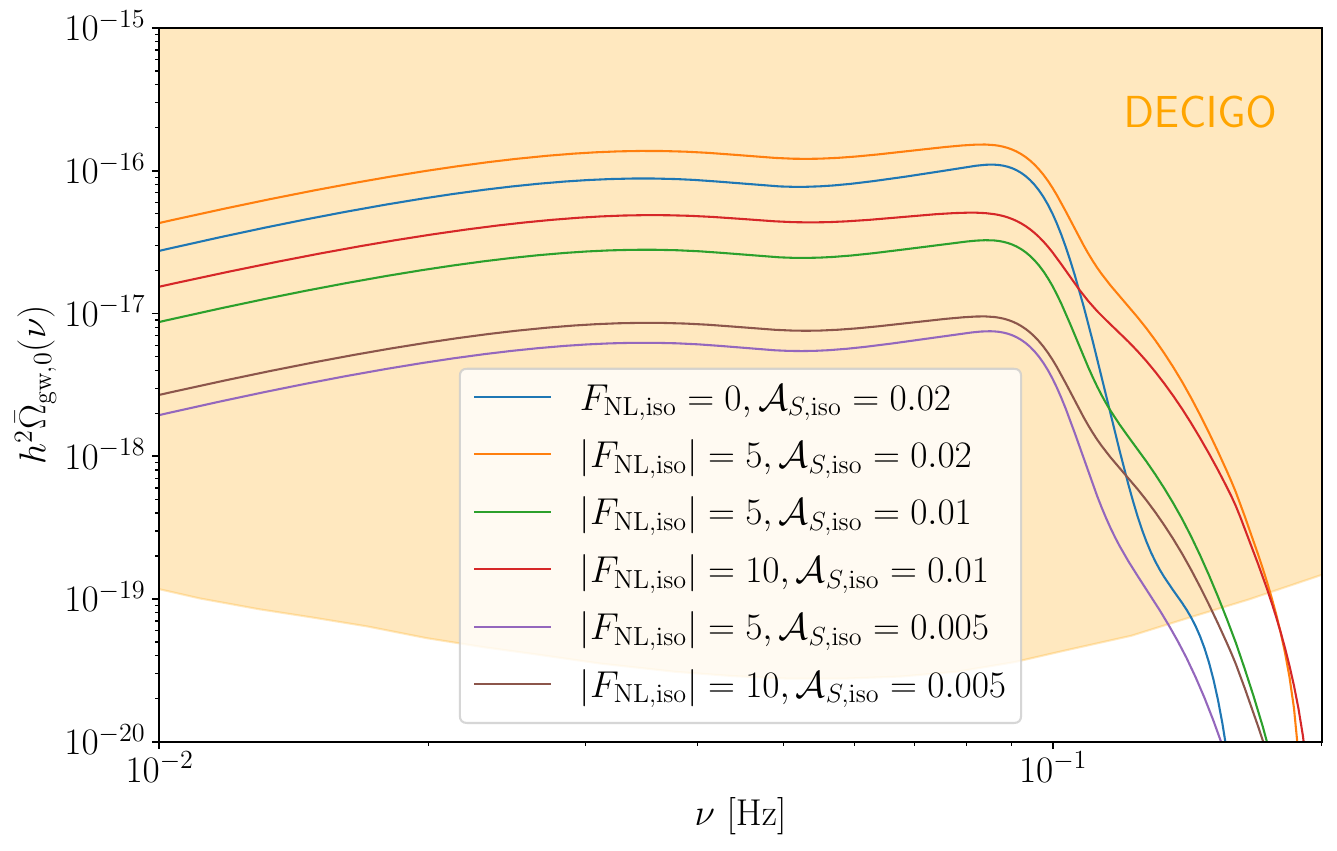}
    \caption{The present-day physical energy-density fraction spectrum $h^2\bar{\Omega}_{\rm gw,0}$ as a function of the \ac{GW} frequency $\nu$ for different values of $|\Fnl|$ and $\cA_{S,\mathrm{iso}}$.
    We compare $h^2\bar{\Omega}_{\rm gw,0}$ with the sensitivity region of DECIGO (orange shaded area) \cite{Seto:2001qf,Kawamura:2020pcg}.
    Other parameters are set as $k_{\mathrm{uv}}/k_{\mathrm{eeq}}=50$ and $\nu_{\mathrm{uv}}= k_\mathrm{uv}/(2\pi)=0.1\,$Hz.}
    \label{fig:Omega_0}
\end{figure*}

Regarding observations, we consider the present-day physical energy-density fraction spectrum, denoted as $h^2\bar{\Omega}_{\rm gw,0}(\nu)$, i.e,
\begin{align}\label{eq:spectrumdensitytoday2}
	h^2\bar{\Omega}_{\rm gw,0}(\nu)\simeq h^2\Omega_{\rm rad,0}
	\bar{\Omega}_\mathrm{gw,e}(k) \,
\Big| _{k=2\pi \nu}\ ,
\end{align}
where $\nu$ denotes the \ac{GW} frequency, $h^2\Omega_{\rm rad,0}\simeq4.18\times 10^{-5}$ is the present-day physical density fraction of radiation with $h$ being the dimensionless Hubble constant \cite{Planck:2018vyg}.
The explicit expression in Eq.~(\ref{eq:spectrumdensitytoday2}) may be influenced by the following two factors.
Firstly, the change in the effective number of relativistic degrees of freedom in the Universe contributes a factor of $\left(g_{*\rho,\rm e}/g_{*\rho,0}\right)\left(g_{*s,\rm e}/g_{*s,0}\right)^{-4/3}$, where $g_{*\rho}$ (or $g_{*s}$) represents the effective number of relativistic degrees of freedom in the energy density (or entropy). 
This factor is model-dependent and is anticipated to be of $\cO(1)$.
Secondly, the decay of solitons into relativistic particles could introduce additional radiation components into the Universe, potentially affecting the result in Eq.~(\ref{eq:spectrumdensitytoday2}).
Since we have assumed that the solitons decay at $\eta\simeq\eta_\mathrm{eeq}$ in order to avoid the soliton domination, this effect is expected to contribute only a correction factor of $\cO(1)$. 
We plot Fig.~\ref{fig:Omega_0} to illustrate the frequency dependence of $h^2\bar{\Omega}_{\uGW, 0}(\nu)$ for different values of $|\Fnl|$ and $\cA_{S,\mathrm{iso}}$. 
It is shown that $\bar{\Omega}_{\uGW, 0} $ exhibits a double-peak structure at $\nu\lesssim \nu_\mathrm{uv}$ and sharply declines at $\nu\gtrsim \nu_\mathrm{uv}$.
The amplitude of $\bar{\Omega}_{\uGW, 0} $ is mainly determined by $\cA_{S,\mathrm{iso}}$ with $\bar{\Omega}_{\uGW, 0} \propto \cA_{S,\mathrm{iso}}^2$.
For $\cO(0.1)\lesssim\cA_{S,\mathrm{iso}} \Fnl^2 <1$, $\bar{\Omega}_{\uGW, 0} $ could also be enhanced due to the contribution of non-Gaussianity.
Moreover, these \ac{GW} signals could fall in the frequency band of \ac{GW} detectors like \ac{DECIGO} \cite{Seto:2001qf,Kawamura:2020pcg} (see Refs.~\cite{Lozanov:2023aez,Lozanov:2023knf} for specific scenarios).
In Fig.~\ref{fig:Omega_0}, we also compare $h^2\bar{\Omega}_{\uGW, 0}(\nu)$ with the sensitivity curve of \ac{DECIGO}.
The isocurvature-induced \acp{GW} involved in our work are potentially detectable for \ac{DECIGO} in future.

Our findings regarding the effects of non-Gaussianity on $\bar{\Omega}_\uGW$ of isocurvature-induced \acp{GW} have the following implications.
Firstly, we demonstrate that the shape of $\bar{\Omega}_\mathrm{gw,0}$ for the ``universal \acp{GW}" from solitons, as discussed in Ref.~\cite{Lozanov:2023aez}, remains nearly unchanged in the perturbative regime, regardless of the presence of non-Gaussianity in soliton isocurvature perturbations.
Consequently, our research broadens the applicable condition for such ``universal \acp{GW}" as a distinct signal for detecting solitons.
Secondly, since the effects of non-Gaussianity on $\bar{\Omega}_\mathrm{gw,0}$ are highly degenerate, relying solely on $\bar{\Omega}_\mathrm{gw,0}$ for measurements of non-Gaussianity is unlikely to be effective.
Therefore, it is quite imperative to establish novel observables in the isocurvature-induced \acp{GW} to detect the non-Gaussianity. 
This provides strong motivations for our subsequent exploration of the anisotropies in the isocurvature-induced \acp{GW}.

\section{Anisotropies in the isocurvature-induced GWs}
\label{sec:aps}

In this section, we move on to study the inhomogeneities in the energy density of isocurvature-induced \acp{GW}, which are mapping as the anisotropies over the skymap. 
The non-Gaussianity indicates the interaction between large- and small-scale isocurvature perturbations, which can redistribute the energy density of \acp{GW} and lead to inhomogeneities at superhorizon scales. 
These inhomogeneities, combined with the propagation effects of \acp{GW} \cite{Contaldi:2016koz,Bartolo:2019oiq,Bartolo:2019yeu,Schulze:2023ich}, ultimately give rise to the observed anisotropies of \acp{GW}.
We will obtain the (reduced) angular power spectrum as an observable characterizing these anisotropies.

\subsection{Reduced angular power spectrum}

The anisotropies in isocurvature-induced \acp{GW} are related to the density contrast of \acp{GW} at the emission time, namely, $\delta_\mathrm{gw,e} (\bk)=\delta_\mathrm{gw} (\eta_\mathrm{e},\bx_\mathrm{e},\bk)$.
Since the angle subtended by the horizon at $\eta_\mathrm{e}$ is extremely small compared to the angular resolution of \ac{GW} detectors, the \ac{GW} signal along the line of sight is actually an average over a large number of such horizons.
The small-scale $\delta_\mathrm{gw,e}$ would be averaged to zero, and thus makes no contribution to observed anisotropies. 
Therefore, the anisotropies of \acp{GW} originate solely from large-scale $\delta_\mathrm{gw,e}$.
This large-scale $\delta_\mathrm{gw,e}$ could be generated by the non-Gaussianity in $\cS$, which introduces the coupling between $\cS_{gS}$ and $\cS_{gL}$.
As aforementioned, the subscripts $_{S}$ and $_{L}$, respectively, stand for short- and long-wavelength modes. 
The \acp{GW} induced by $\cS_{S}$ can be spatially modulated at large scales by $\cS_{gL}$, through the non-linear term $\cS_{S}\sim\Fnl \cS_{gS} \cS_{gL}$ in Eq.~(\ref{eq:fnl-def-k}). 
To illustrate the above picture, we expand $\rho_\mathrm{gw}\sim\langle\cS^4\rangle$ to the linear order of $\cS_{gL}$, i.e.,
\begin{equation}\label{eq:expansion}
    \langle\cS^4\rangle
    =\langle\cS_S^4\rangle+
    \cO(\cS_{gL})\,\Fnl \langle \cS_S^3\, \cS_{gS}\rangle
    +\cO(\cS^2_{gL})\ .
\end{equation}
On the right-hand side of Eq.~(\ref{eq:expansion}), the first term is corresponding to $\bar{\Omega}_\mathrm{gw,e}$, 
while the second term gives the leading order of the large-scale density contrast due to the spatial modulation of $\cS_{gL}$, namely $\delta_\mathrm{gw,e}\sim \Fnl \langle \cS_S^3\, \cS_{gS}\rangle\, \cS_{gL}$.
Employing the diagrammatic approach akin to the study of \acp{SIGW} \cite{Bartolo:2019zvb,Li:2023qua,Li:2023xtl}, the explicit expression of $\delta_\mathrm{gw,e}$ is showed as
\begin{equation}\label{eq:deltaGW}
    \delta_\mathrm{gw,e}(\bk)
    =
    \Fnl\,
    \frac{\bar{\Omega}_{\mathrm{ng,e}} (k)}{\bar{\Omega}_\mathrm{gw,e} (k)}
    \int \frac{\ud^{3}\bq}{(2\pi)^{3/2}} e^{i\bq\cdot\bx_\mathrm{e}} 
    \cS_{gL}(\bq)
    \ ,
\end{equation}
where $\bar{\Omega}_{\mathrm{ng,e}}$ is defined as
\begin{equation}
    \bar{\Omega}_{\mathrm{ng,e}} (k) 
    = 2^3 \bar{\Omega}_\mathrm{gw,e}^{G} (k)  
        + 2^2 \bar{\Omega}_\mathrm{gw,e}^H (k)  
        + 2^2 \bar{\Omega}_\mathrm{gw,e}^C (k) 
        + 2^2 \bar{\Omega}_\mathrm{gw,e}^Z (k) \ .
\end{equation}
Additionally, the $\cO(\cS^2_{gL})$ term in Eq.~(\ref{eq:expansion}) can be safely neglected when considering $\cA_{L,\mathrm{iso}}\ll \cA_{S,\mathrm{iso}}$ in our present work.

In order to get the present-day density contrast $\delta_\mathrm{gw,0}(\bk)=\delta_\mathrm{gw}(\eta_\mathrm{0},\bx_\mathrm{0},\bk)$ for an observer located at $(\eta_0,\bx_0)$, we solve the Boltzmann equation governing the evolution of \acp{GW}, following a line-of-sight approach \cite{Contaldi:2016koz,Bartolo:2019oiq,Bartolo:2019yeu}.
The $\delta_{\uGW,0}$ consists of $\delta_\mathrm{gw,e}$ and propagation effects in a way that is similar to the study of \acp{SIGW} \cite{Bartolo:2019zvb,Li:2023qua,Li:2023xtl,Rey:2024giu,Ruiz:2024weh,Schulze:2023ich,LISACosmologyWorkingGroup:2022kbp,LISACosmologyWorkingGroup:2022jok,Malhotra:2022ply,Wang:2023ost,Yu:2023jrs}.
It is given by 
\begin{equation}\label{eq:delta-0}
    \delta_{\uGW,0}(\bk) = \delta_\mathrm{gw,e} (\bk) + \left[4-n_\mathrm{gw,0} (k)\right] \Phi (\eta_\mathrm{e}, \bx_\mathrm{e})+...\ ,
\end{equation}
where the second term on the right-hand side arises from gravitational redshift or blueshift caused by cosmological scalar perturbations, known as the \ac{SW} effect \cite{Sachs:1967er}, and the ellipsis could include other propagation effects such as the \ac{ISW} effect \cite{Sachs:1967er}, which is usually expected to be negligible (e.g., see Ref.~\cite{Bartolo:2019zvb} for the case of \acp{SIGW}). 
Nonetheless, we can take these effects into account if necessary \cite{Schulze:2023ich,Cai:2024dya,Zhao:2024gan}. 
In Eq.~(\ref{eq:delta-0}), the spectral index $n_{\uGW,0} (k)$ is defined as
\begin{equation}\label{eq:ngw}
    n_\mathrm{gw,0} (k) = \frac{\partial \ln \bar{\Omega}_\mathrm{gw,0} (\nu)}{\partial\ln \nu}\bigg| _{k=2\pi \nu} \simeq  \frac{\partial \ln \bar{\Omega}_\mathrm{gw,e} (k)}{\partial\ln k}\ ,
\end{equation}
and the large-scale scalar perturbation $\Phi (\eta_\mathrm{e}, \bx_\mathrm{e})$ is approximately given by large-scale curvature perturbations, i.e.,
\begin{equation}\label{eq:SWe}
    \Phi (\eta_\mathrm{e}, \bx_\mathrm{e}) \simeq \frac{3}{5} \int \frac{\ud^{3}\bq}{(2\pi)^{3/2}} e^{i\bq\cdot\bx_\mathrm{e}} \zeta_L(\bq)\ ,
\end{equation}
where we neglect contributions of large-scale isocurvature perturbations due to $\cA_{L,\mathrm{iso}}\ll\cA_{\mathrm{ad}}$.

Assuming the cosmological principle, we define the reduced angular power spectrum to describe the anisotropies of \acp{GW}, i.e.,
\begin{equation}\label{eq:Ctilde-def}
    \langle\delta_{\uGW,0,\ell_1 m_1}(k_1) \delta_{\uGW,0,\ell_2 m_2}^\ast(k_2)\rangle
    = \delta_{\ell_1 \ell_2} \delta_{m_1 m_2} \tilde{C}_\ell (k_1, k_2)\ ,
\end{equation}
which can be evaluated within the same frequency band (i.e., $k_1=k_2$) or across different frequency bands (i.e., $k_1\neq k_2$). 
Here, $\delta_{\uGW,0,\ell m}$ denotes the coefficient in the spherical harmonic expansion of $\delta_{\text{gw},0}$, namely,
\begin{equation}\label{eq:shsai}
    \delta_{\uGW,0}(\bk) = \sum_{\ell m} \delta_{\uGW,0,\ell m}(k) Y_{\ell m}(\hat{\bk})\ .
\end{equation}
Analogous to calculations in Refs.~\cite{Bartolo:2019zvb,Bartolo:2019oiq,Bartolo:2019yeu,Li:2023qua,Li:2023xtl,Schulze:2023ich,Rey:2024giu,Ruiz:2024weh}, the reduced angular power spectrum of isocurvature-induced \acp{GW} can be derived from Eqs.~(\ref{eq:deltaGW}-\ref{eq:shsai}).
It is given by
\begin{equation}\label{eq:reduced-APS}
    \tilde{C}_\ell (k_1, k_2) 
    = \frac{2\pi \cA_{\mathrm{ad}}}{\ell (\ell+1)} 
        \left\{
            \beta\Fnl^2
            \frac{\bar{\Omega}_{\mathrm{ng,e}} (k_1)}{\bar{\Omega}_{\mathrm{gw,e}} (k_1)}
            \frac{\bar{\Omega}_{\mathrm{ng,e}} (k_2)}{\bar{\Omega}_{\mathrm{gw,e}} (k_2)}
            + \frac{9}{25} \left[4 - n_{\uGW,0} (k_1)\right] \left[4 - n_{\uGW,0} (k_2)\right]
        \right\}\ ,
\end{equation}
where we use the large-scale correlations, i.e., $\langle \zeta_{L} \zeta_{L} \rangle \sim \cA_{\mathrm{ad}}$, $\langle \cS_{gL} \cS_{gL} \rangle \sim \cA_{L,\mathrm{iso}}=\beta \cA_{\mathrm{ad}}$, and $\langle \zeta_{L} \cS_{gL} \rangle =0$.
The first and second terms in the bracket of Eq.~\eqref{eq:reduced-APS} correspond to $\delta_\mathrm{gw,e}$ and the \ac{SW} effect term, respectively.

Eq.~(\ref{eq:reduced-APS}) is one of the key results of our work.
It is essential to compare Eq.~(\ref{eq:reduced-APS}) with the reduced angular power spectrum of \acp{SIGW} (e.g., see Eq.~(5.16) in Ref.~\cite{Li:2023qua}), with the main differences emphasized as follows.
Firstly, the expression for $\tilde{C}_\ell$ in Eq.~(\ref{eq:reduced-APS}) introduces a novel parameter $\beta$. 
It is easily understood, because two $\delta_\mathrm{gw,0}$ with large spatial separation are correlated by $\cS_{gL}$ for the case of isocurvature-induced \acp{GW}, while by $\zeta_L$ for the case of \acp{SIGW}.
Secondly, in Eq.~(\ref{eq:reduced-APS}), the cross term between $\delta_\mathrm{gw,e}$ and the \ac{SW} effect does not exist, different from the case of \acp{SIGW}.
This is because we have $\delta_\mathrm{gw,e}\propto\cS_{gL}$ and $(4-n_{\uGW}) \Phi \propto \zeta_L$ in Eq.~(\ref{eq:delta-0}), leading to the vanishing correlation between them, i.e., $\langle \zeta_{L} \cS_{gL} \rangle =0$.
This results in some unique features in the anisotropies of isocurvature-induced \acp{GW}, as will be demonstrated below.

\subsection{Results}

\begin{figure*}[htbp]
    \centering
    \includegraphics[width=0.8\textwidth]{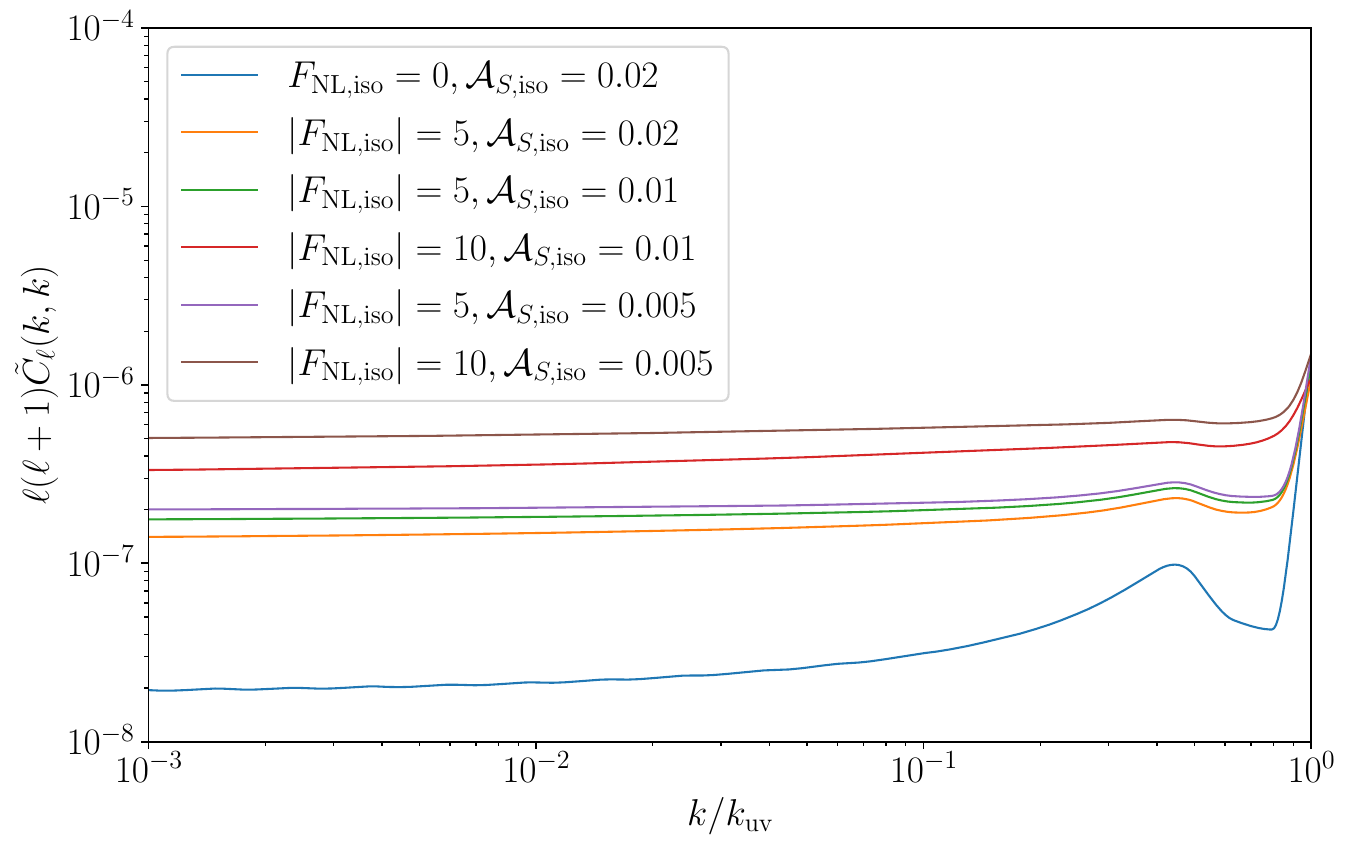}
    \caption{Reduced angular power spectrum $\ell(\ell+1)\Tilde{C}_\ell$ versus the \ac{GW} wavenumber $k$ for various values of $|\Fnl|$ and $\cA_{S,\mathrm{iso}}$.
    We set $\beta=10^{-2}$.}
    \label{fig:reduced-C_l}
\end{figure*}

In Fig.~\ref{fig:reduced-C_l}, we illustrate the $k$-dependence of $\ell(\ell+1)\Tilde{C}_\ell(k,k)$ for various values of $|\Fnl|$ and $\cA_{S,\mathrm{iso}}$.
It is illustrated that $\ell(\ell+1)\Tilde{C}_\ell$ is primarily influenced by $|\Fnl|$ but shows little sensitivity to $\cA_{S,\mathrm{iso}}$.
The anisotropies can be boosted by approximately two orders of magnitude due to non-Gaussianity, as compared to the Gaussian cases.
Therefore, the detection of the anisotropies could be helpful to break the high degeneracy of $|\Fnl|$ on $\bar{\Omega}_\mathrm{gw}$.
We note that the non-linearity parameter $\Fnl$ exhibits a strict sign degeneracy in $\ell(\ell+1)\Tilde{C}_\ell$ due to the absence of the cross term between $\delta_\mathrm{gw,e}$ and the \ac{SW} effect in Eq.~\eqref{eq:reduced-APS}, different from the case of \acp{SIGW}.
Additionally, the sharp increase in $\ell(\ell+1)\Tilde{C}_\ell$ at $k\simeq k_{\rm uv}$ is attributed to the pronounced \ac{SW} effect, which arises from the rapid decrease in $\bar{\Omega}_\uGW$ at $k\simeq k_{\rm uv}$ due to the unphysical cutoff in $\Delta^2_{\cS_g}$.

\begin{figure*}[htbp]
    \centering
    \includegraphics[width=0.8\textwidth]{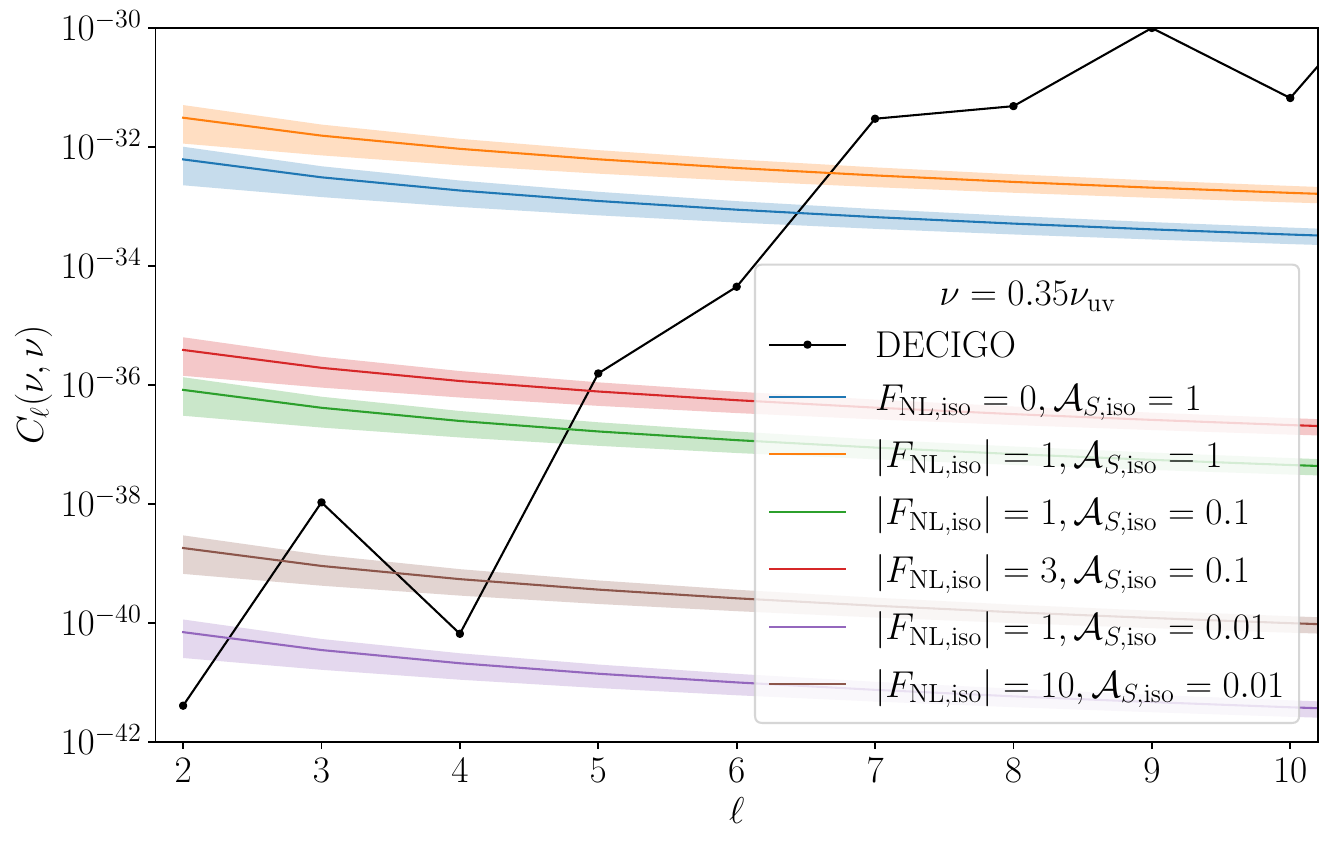}
    \caption{Multipole dependence of the angular power spectrum $C_\ell(\nu,\nu)$ for different values of $|\Fnl|$ and $\cA_{S,\mathrm{iso}}$.
    Shaded regions represent the uncertainties (68\% confidence level) due to the cosmic variance, i.e., $\Delta C_{\ell}/C_{\ell}=\sqrt{2/(2\ell+1)}$.
    We compare $C_\ell(\nu,\nu)$ to the noise angular power spectrum of \ac{DECIGO} \cite{Braglia:2021fxn}, with a frequency $\nu$ aligning with the optimal sensitivity frequency.
    Other parameters are set as $\beta=10^{-2}$ and $k_{\mathrm{uv}}/k_{\mathrm{eeq}}=20$.}
    \label{fig:C_l}
\end{figure*}

Multiplying by the energy-density fraction spectrum, the reduced angular power spectrum is further reformulated as the angular power spectrum, i.e.,
\begin{equation}\label{eq:cellsai}
C_{\ell}(\nu_1, \nu_2) =
    \frac{\bar{\Omega}_{\uGW,0}(\nu_1)}{4\pi}
    \frac{\bar{\Omega}_{\uGW,0}(\nu_2)}{4\pi}
    \tilde{C}_{\ell}(k_1, k_2)\,\Big|_{k_1=2\pi\nu_1,\, k_2=2\pi\nu_2}\ .
\end{equation}
In Fig.~\ref{fig:C_l}, we depict the multipole dependence of $C_\ell(\nu,\nu)$ for different values of $|\Fnl|$ and $\cA_{S,\mathrm{iso}}$, with $\nu = 0.35\,\nu_{\rm uv}$ positioned close to the left peak of $\bar{\Omega}_\mathrm{gw}$.
Similar to the multipole dependence of ${C}_\ell$ of \acp{SIGW}, the multipole dependence of ${C}_\ell$ of isocurvature-induced \acp{GW} also follows ${C}_\ell \propto [\ell(\ell+1)]^{-1}$, which is instrumental in differentiating them from other sources of \acp{GW} and astrophysical foregrounds.
Increasing $|\Fnl|$ and $\cA_{S,\mathrm{iso}}$ could lead to an increase of ${C}_\ell$.
Comparing $C_\ell(\nu,\nu)$ with the optimal sensitivity of \ac{DECIGO}  \cite{Braglia:2021fxn}, as shown in Fig.~\ref{fig:C_l}, we infer that only for large values of $\cA_{S,\mathrm{iso}}$, e.g., $\cA_{S,\mathrm{iso}}\gtrsim \cO(0.1)$, the anisotropies of isocurvature-induced \acp{GW} are potentially observed at low-$\ell$ multipoles by \ac{DECIGO}.

\begin{figure*}[htbp]
    \centering
    \includegraphics[width=1\textwidth]{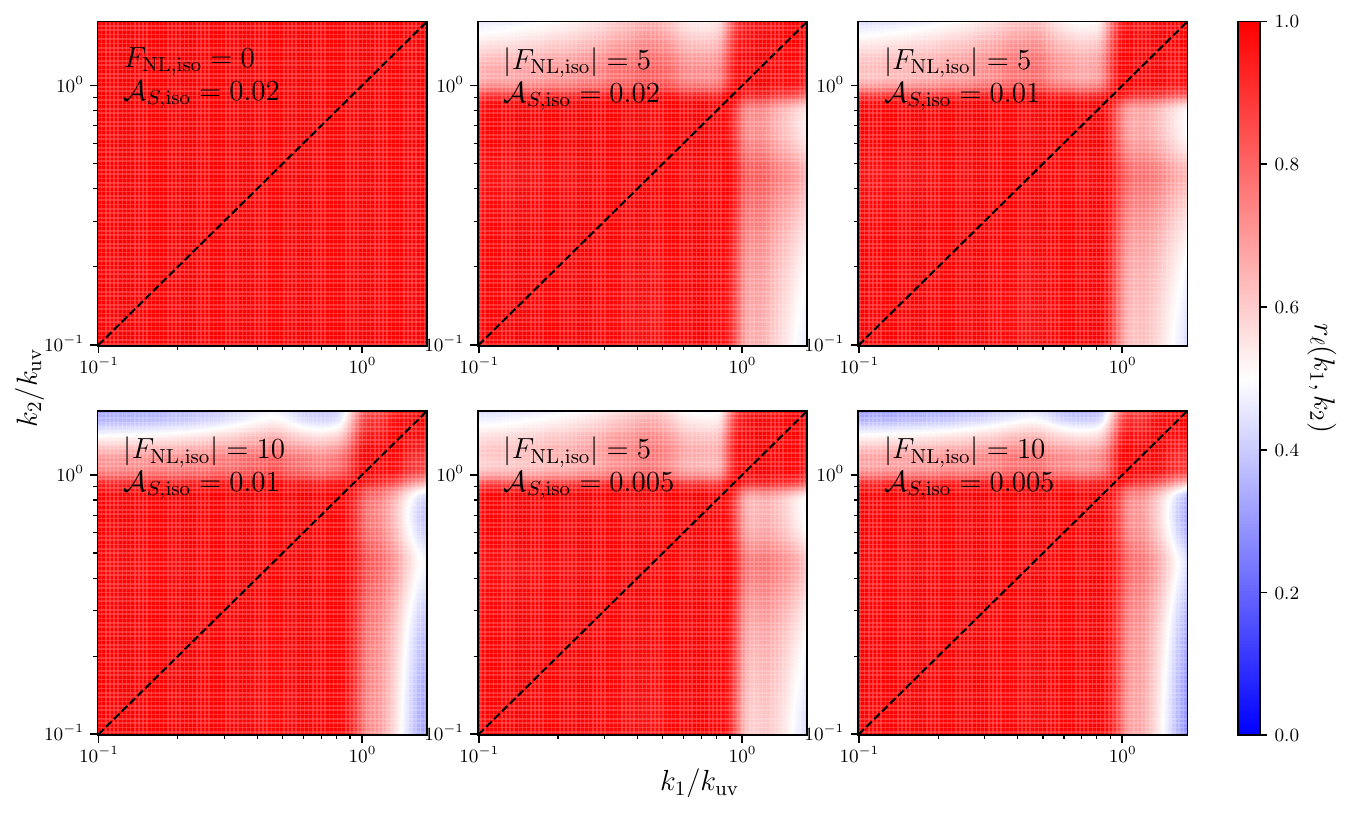}
    \caption{Cross-correlation factor $r_\ell(k_1,k_2)$ with respect to the \ac{GW} wavenumber band for different values of $|\Fnl|$ and $\cA_{S,\mathrm{iso}}$. 
    In each panel, the dashed line refers to $k_1=k_2$.
    We use the same sets of model parameters as those of Fig.~\ref{fig:Omega_0} and set $\beta=10^{-2}$ for all cases.}
    \label{fig:r_l}
\end{figure*}

In contrast to the study of the \ac{CMB}, for which only the correlation between the same frequency band is considered due to the blackbody nature of the \ac{CMB}, we define the cross-correlation between different frequency bands of isocurvature-induced \acp{GW} as 
\begin{equation}\label{eq:cfactor-def}
r_{\ell} (k_1,k_2)=
\frac{\tilde{C}_{\ell}(k_1,k_2)}{\sqrt{\tilde{C}_{\ell}(k_1,k_1)\tilde{C}_{\ell}(k_2,k_2)}}\ .
\end{equation}
In Fig.~\ref{fig:r_l}, we visualize it for different values of $|\Fnl|$ and $\cA_{S,\mathrm{iso}}$.
In the regimes of ($k_1\lesssim k_\mathrm{uv}, k_2\lesssim k_\mathrm{uv}$) and ($k_1\gtrsim k_\mathrm{uv}, k_2\gtrsim k_\mathrm{uv}$), we have $r_\ell(k_1,k_2)\simeq 1$.
However, in the regimes of ($k_1\gtrsim k_\mathrm{uv}, k_2\lesssim k_\mathrm{uv}$) and ($k_1\lesssim k_\mathrm{uv}, k_2\gtrsim k_\mathrm{uv}$), we find $r_\ell(k_1,k_2)<1$, indicating that the anisotropies for different frequency bands are not fully correlated.
Our analysis also indicates that $r_\ell(k_1,k_2)$ can be affected by $|\Fnl|$.
In the case of Gaussianity, we consistently have $r_\ell(k_1,k_2)= 1$.
However, $r_\ell(k_1,k_2)$ decreases as $|\Fnl|$ increases in the regimes of ($k_1\gtrsim k_\mathrm{uv}, k_2\lesssim k_\mathrm{uv}$) and ($k_1\lesssim k_\mathrm{uv}, k_2\gtrsim k_\mathrm{uv}$).
Therefore, $r_\ell(k_1,k_2)$ could also be a probe to discern the value of $|\Fnl|$.

\section{Conclusions and discussion}
\label{sec:conc}

In this study, we have investigated the \acp{GW} induced by the non-Gaussian soliton isocurvature perturbations, encompassing both their isotropic background and anisotropies.
As an important result for the isotropic background, we found that the non-Gaussianity would not significantly affect both the amplitude and shape of the energy-density fraction spectrum in the perturbative regime.
This result indicates that the ``universal \acp{GW}'' still serves as a unique signal to identify solitons, even if the soliton isocurvature perturbations are non-Gaussian.
The insensitive dependence of the energy-density fraction spectrum on $|\Fnl|$ could bring a significant challenge to measure the non-Gaussianity.
To address this challenge, we provided the first analysis of the anisotropies of isocurvature-induced \acp{GW}, accounting for both $\delta_\mathrm{gw,e}$ and the propagation effects.
As another key result in this work, the (reduced) angular power spectrum, as demonstrated by Eq.~\eqref{eq:reduced-APS}, could serve as a powerful probe of the non-Gaussianity in isocurvature perturbations. 
It highlights the significant role of this non-Gaussianity in enhancing anisotropies in isocurvature-induced \acp{GW}, breaking the degeneracies between $|\Fnl|$ and other model parameters such as $\mathcal{A}_{S,\mathrm{iso}}$.
The expression of the (reduced) angular power spectrum also showcases the intrinsic differences between the anisotropies of isocurvature-induced \acp{GW} and those of \acp{SIGW}.
The isotropic background and anisotropies of isocurvature-induced \acp{GW} can provide novel probes and signatures for diverse cosmological models related to soliton formation, enriching our understanding of the early Universe.

As a related work, Ref.~\cite{Domenech:2021and} discussed the effects of non-Gaussianity on the \acp{GW} induced by dark matter isocurvature perturbations.
The authors estimate that the non-Gaussianity may significantly enhance $\Omega_\mathrm{gw,0}$, which seems different from our result.
However, these two are not contradictory in essence.
In fact, the differences mainly come from the following two aspects.
(a) In Ref.~\cite{Domenech:2021and}, \acp{GW} are produced in standard radiation-matter Universe with a suppression of power-law $(k/k_\mathrm{eq})^{-4}$ on $\bar{\Omega}_\mathrm{gw,e}(k)$, where $k_\mathrm{eq}\sim 0.01\, \mathrm{Mpc}^{-1}$ is the mode reentering the horizon at standard radiation-matter equality.
To generate detectable \acp{GW} within the frequency band of \ac{GW} detectors, $\cA_{S,\mathrm{iso}}$ needs to be much larger than unity to overcome the suppression factor.
However, in our paper, the corresponding suppression factor is $(k/k_\mathrm{eeq})^{-4}$ for an early radiation-soliton Universe.
We consider the case $\cA_{S,\mathrm{iso}} \lesssim 1$, which is enough to generate detectable \acp{GW}.
(b) In Ref.~\cite{Domenech:2021and}, the \ac{GW} sources are highly skewed non-Gaussian $\cS$ that can not be parameterized as Eq.~(\ref{eq:fnl-def-k}).
Whereas in our present work, we consider the non-Gaussian $\cS$ in the form of Eq.~(\ref{eq:fnl-def-k}).

In this work, we have assumed that solitons decay before they dominate the early Universe. 
If solitons could dominate the early Universe, their decay might introduce additional relativistic degrees of freedom and also change the energy density of radiation, depending on specific models and the duration of soliton domination.
These effects would notably suppress the amplitudes of $\bar{\Omega}_{\rm gw,0}$ and $C_\ell$, making them more challenging to be observed.
However, other main results presented in this paper are expected to remain unchanged.

We may further investigate cross-correlations between the isocurvature-induced \acp{GW} with the \ac{CMB} temperature anisotropies and polarization. 
Recent studies have explored the cross-correlations between \acp{SIGW} and the \ac{CMB} \cite{Dimastrogiovanni:2022eir,Schulze:2023ich,Zhao:2024gan}.
The cross-correlations could be significant since the anisotropies of \acp{SIGW} and the \ac{CMB} share the same origin, i.e., curvature perturbations.
In contrast, the isocurvature-induced \acp{GW} exhibit minimal cross-correlations with the \ac{CMB}, since their anisotropies mainly originate from the isocurvature perturbations (with other effects such as the \ac{SW} effect being subdominant), which are expected not to correlate with the curvature perturbations.
These distinctions reveal that the cross-correlations between the \ac{GW} signal and the \ac{CMB} could serve as crucial observable to identify the isocurvature-induced \acp{GW}.

Our study of isocurvature-induced \acp{GW} may provide important implications for \acp{PBH}, which are a promising candidate for dark matter.
If solitons temporarily dominate the early Universe before their decay, the overdensities caused by their number density fluctuations could be the seeds for \ac{PBH} formation \cite{Cotner:2016cvr,Cotner:2017tir,Cotner:2018vug,Cotner:2019ykd,Flores:2021jas}, accompanied by \acp{GW} induced by soliton isocurvature perturbations. 
For comparison, \acp{PBH} could also be produced through the gravitational collapse of enhanced small-scale curvature perturbations \cite{Hawking:1971ei}, accompanied by the production of \acp{SIGW}.
The above two mechanisms of \ac{PBH} formation are potentially distinguishable via identifying the corresponding \ac{GW} signals, e.g., through their cross-correlations with the \acp{CMB}.
Furthermore, the presence of non-Gaussianity in soliton isocurvature perturbations may modify the abundance of \acp{PBH} and also lead to a clustering of \acp{PBH}. 
However, this intriguing topic is beyond the scope of our current study. Hence, we leave it to future works. 

Our present work was focused on the case that the soliton isocurvature perturbations and curvature perturbations are initially statistical independent, i.e., $\langle\zeta(\eta_\mathrm{i},\bk_1)\cS(\eta_\mathrm{i},\bk_2)\rangle=0$.
However, the correlation between them may exist if they are coupled during the inflation (e.g., see Refs.~\cite{Tsujikawa:2002qx,Byrnes:2006fr,Lalak:2007vi}) or during soliton formation scenarios (e.g., \acp{PBH} from enhanced curvature perturbations).
The non-vanishing correlation $\langle\zeta(\eta_\mathrm{i},\bk_1)\cS(\eta_\mathrm{i},\bk_2)\rangle$, which is likely to be highly model-dependent, could alter the anticipated anisotropies of isocurvature-induced \acp{GW}, as well as the cross-correlation between the isocurvature-induced \acp{GW} and the \ac{CMB}. 
We defer a comprehensive study of this topic to future works.

\acknowledgments
S.W. would like to express his gratitude to Xian Gao and Jiarui Sun for their warm hospitality during the final stage of this work, which was conducted during a visit to the Sun Yat-sen University. 
This work is supported by the National Key R\&D Program of China No. 2023YFC2206403 and the National Natural Science Foundation of China (Grant No. 12175243).

\appendix

\section{Formulas for the energy-density fraction spectrum}
\label{sec:inte}

In this Appendix, we summarize the formulas for $\bar{\Omega}_\mathrm{gw,e}^X$ ($X=G,H,C,Z,R,P,N$) in Eq.~(\ref{eq:OmegaX}) as forms suitable for numerical integration. Similar results for the case of \acp{SIGW} can be found in Refs.~\cite{Adshead:2021hnm,Ragavendra:2021qdu,Abe:2022xur,Li:2023qua,Li:2023xtl,Yuan:2023ofl,Perna:2024ehx,Ruiz:2024weh}.
To be specific, the $\bar{\Omega}_\mathrm{gw,e} ^{(0)}$ component is given by of $\bar{\Omega}^G_\mathrm{gw,e}$, namely,
\begin{eqnarray}\label{eq:A1}
    \bar{\Omega}_\mathrm{gw,e}^G (k) 
    &=& \frac{1}{3} \int_0^\infty \ud t_1 \int_{-1}^1 \ud s_1 
        \overbar{J^2 (u_1,v_1,\kappa, x\rightarrow\infty)} \frac{1}{(u_1 v_1)^2} 
        \Delta^2_{\cS_g} (v_1 k) \Delta^2_{\cS_g} (u_1 k) \ .\label{eq:Omega-G}
\end{eqnarray}
The $\bar{\Omega}_\mathrm{gw,e} ^{(1)}$ component consists of $\bar{\Omega}^H_\mathrm{gw,e}$, $\bar{\Omega}^C_\mathrm{gw,e}$, and $\bar{\Omega}^Z_\mathrm{gw,e}$, namely,
\begin{eqnarray}
    \bar{\Omega}_\mathrm{gw,e}^H (k)
    &=& \frac{1}{3} \Fnl^2 
        \prod_{i=1}^2 \biggl[\int_0^\infty \ud t_i \int_{-1}^1 \ud s_i\biggr] 
        \overbar{J^2 (u_1,v_1,\kappa,x\rightarrow\infty)}\frac{1}{(u_1 v_1 u_2 v_2)^2}\nonumber\\ 
    & &\hphantom{\Fnl^2 \frac{1}{12} \prod_{i=1}^2}
        \times \Delta^2_{\cS_g} (v_1 v_2 k) \Delta^2_{\cS_g} (u_1 k) \Delta^2_{\cS_g} (v_1 u_2 k) \ ,\label{eq:Omega-H}\\
        \nonumber \\ 
        \bar{\Omega}_\mathrm{gw,e}^C (k) 
    &=& \frac{1}{3\pi} \Fnl^2 
        \prod_{i=1}^2 \biggl[\int_0^\infty \ud t_i \int_{-1}^1 \ud s_i\, v_i u_i\biggr] 
        \int_0^{2\pi} \ud \varphi_{12}\, \cos 2\varphi_{12}  \nonumber\\
        & &\hphantom{\Fnl^2 \frac{1}{12\pi} \prod_{i=1}^2 }
        \times \overbar{J (u_1,v_1,\kappa,x\rightarrow\infty)J (u_2,v_2,\kappa,x\rightarrow\infty)} \nonumber\\
        & &\hphantom{\Fnl^2 \frac{1}{12\pi} \prod_{i=1}^2 }
        \times \frac{\Delta^2_{\cS_g} (v_2 k)}{v_2^3} \frac{\Delta^2_{\cS_g} (u_2 k)}{u_2^3} \frac{\Delta^2_{\cS_g} (w_{12} k)}{w_{12}^3} 
            \ ,\label{eq:Omega-C}\\
            \nonumber \\ 
    \bar{\Omega}_\mathrm{gw,e}^Z (k) 
    &=& \frac{1}{3\pi} \Fnl^2 
        \prod_{i=1}^2 \biggl[\int_0^\infty \ud t_i \int_{-1}^1 \ud s_i\, v_i u_i\biggr] 
        \int_0^{2\pi} \ud \varphi_{12}\, \cos 2\varphi_{12} \nonumber\\
        & &\hphantom{\Fnl^2 \frac{1}{12\pi} \prod_{i=1}^2 }
        \times \overbar{J (u_1,v_1,\kappa,x\rightarrow\infty)J (u_2,v_2,\kappa,x\rightarrow\infty)} \nonumber\\
        & &\hphantom{\Fnl^2 \frac{1}{12\pi} \prod_{i=1}^2 }
        \times \frac{\Delta^2_{\cS_g} (v_2 k)}{v_2^3} \frac{\Delta^2_{\cS_g} (u_1 k)}{u_1^3} \frac{\Delta^2_{\cS_g} (w_{12} k)}{w_{12}^3} \ .
\end{eqnarray}
The $\bar{\Omega}_\mathrm{gw,e} ^{(2)}$ component consists of $\bar{\Omega}^R_\mathrm{gw,e}$, $\bar{\Omega}^P_\mathrm{gw,e}$, and $\bar{\Omega}^N_\mathrm{gw,e}$, namely,
\begin{eqnarray}
    \bar{\Omega}_\mathrm{gw,e}^R (k)
    &=& \frac{1}{12} \Fnl^4 
        \prod_{i=1}^3 \biggl[\int_0^\infty \ud t_i \int_{-1}^1 \ud s_i\biggr] 
        \overbar{J^2 (u_1,v_1,\kappa,x\rightarrow\infty)}\frac{1}{(u_1 v_1 u_2 v_2 u_3 v_3)^2}\nonumber\\ 
    & &\hphantom{\Fnl^2 \frac{1}{12} \prod_{i=1}^2}
        \times \Delta^2_{\cS_g} (v_1 v_2 k) \Delta^2_{\cS_g} (v_1 u_2 k) \Delta^2_{\cS_g} (u_1 v_3 k) \Delta^2_{\cS_g} (u_1 u_3 k) \ ,\\
        \nonumber\\
    \bar{\Omega}_\mathrm{gw,e}^P (k)
    &=& \frac{1}{24\pi^2} \Fnl^4 
        \prod_{i=1}^3 \biggl[\int_0^\infty \ud t_i \int_{-1}^1 \ud s_i\, v_i u_i\biggr] 
        \int_0^{2\pi} \ud \varphi_{12}\ud \varphi_{23}\, \cos 2\varphi_{12} \nonumber\\ 
        & &\hphantom{\Fnl^4 \frac{1}{96\pi^2} \prod_{i=1}^3 }
        \times \overbar{J (u_1,v_1,\kappa,x\rightarrow\infty) J (u_2,v_2,\kappa,x\rightarrow\infty)} \nonumber\\ 
        & &\hphantom{\Fnl^4 \frac{1}{96\pi^2} \prod_{i=1}^3 } 
        \times \frac{\Delta^2_{\cS_g} (v_3 k)}{v_3^3} \frac{\Delta^2_{\cS_g} (u_3 k)}{u_3^3} \frac{\Delta^2_{\cS_g} (w_{13} k)}{w_{13}^3} \frac{\Delta^2_{\cS_g} (w_{23} k)}{w_{23}^3} \ , \\
        \nonumber \\ 
    \bar{\Omega}_\mathrm{gw,e}^N (k) 
    &=& \frac{1}{24\pi^2} \Fnl^4 
        \prod_{i=1}^3 \biggl[\int_0^\infty \ud t_i \int_{-1}^1 \ud s_i\, v_i u_i\biggr] 
        \int_0^{2\pi} \ud \varphi_{12}\ud \varphi_{23}\, \cos 2\varphi_{12} \nonumber\\ 
        & &\hphantom{\Fnl^4 \frac{1}{96\pi^2} \prod_{i=1}^3 }
        \times \overbar{J (u_1,v_1,\kappa,x\rightarrow\infty) J (u_2,v_2,\kappa,x\rightarrow\infty)} \nonumber\\ 
        & &\hphantom{\Fnl^4 \frac{1}{96\pi^2} \prod_{i=1}^3 } 
        \times \frac{\Delta^2_{\cS_g} (u_3 k)}{u_3^3} \frac{\Delta^2_{\cS_g} (w_{13} k)}{w_{13}^3} 
            \frac{\Delta^2_{\cS_g} (w_{23} k)}{w_{23}^3}
            \frac{\Delta^2_{\cS_g} (w_{123} k)}{w_{123}^3} \label{eq:A7} \ .
\end{eqnarray}
In Eqs.~(\ref{eq:A1}-\ref{eq:A7}), we have introduced a new function $J (u_i,v_i,\kappa,x)$, defined as
\begin{equation}\label{eq:J-def}
    J (u_i,v_i,\kappa,x) 
    = \frac{x}{8}\bigl[(v_i+u_i)^2-1\bigr] \bigl[1-(v_i-u_i)^2\bigr] \hat{I} (u_i,v_i,\kappa,x)\ ,
\end{equation}
where we have used $\kappa=k/k_\mathrm{eeq}$ and $x=k\eta$, and $\hat{I} (u_i,v_i,\kappa,x)$ has been given in Eq.~(\ref{eq:I-RD}).
Additionally, we have introduced the following new notations
\begin{align}
    s_i &= u_i - v_i\ ,\\
    t_i &= u_i + v_i -1\ , \\
    \nonumber\\
    y_{ij} &=  \frac{\cos\varphi_{ij}}{4}\sqrt{t_i (t_i + 2) (1 - s_i^2) t_j (t_j + 2) (1 - s_j^2)} 
            + \frac{1}{4}[1 - s_i (t_i + 1)][1 - s_j (t_j + 1)]\ ,\\
    w_{ij} &= \sqrt{v_i^2 + v_j^2 - y_{ij}}\ ,\\
    w_{123} &= \sqrt{v_1^2 + v_2^2 + v_3^2 + y_{12} - y_{13} - y_{23}}\ .
\end{align}








\bibliography{biblio}
\bibliographystyle{JHEP}
\end{document}